\documentclass[aps,prl,notitlepage,amsmath,amssymb,amsfonts,superscriptaddress,twocolumn]{revtex4-2}

\makeatletter
\def\@hangfrom@section#1#2#3{\normalsize\@hangfrom{#1#2}#3}
\def\@hangfroms@section#1#2{\normalsize#1#2}
\makeatother

\usepackage{graphicx,soul}
\usepackage{dcolumn}
\usepackage{bm}
\usepackage{bbm} 

\usepackage{dsfont}
\usepackage{enumerate}

\usepackage[dvipsnames, x11names]{xcolor}
\usepackage[colorlinks=true]{hyperref}
\hypersetup{
    colorlinks=true,
    citecolor=blue,
    linkcolor=red,
    urlcolor=RoyalBlue,
    anchorcolor=Purple
}

\usepackage[percent]{overpic}
\usepackage{tabularx}
\usepackage{multirow}
\usepackage{CJK}
\usepackage{orcidlink}

\usepackage{physics}

\newcommand{\blue}[1]{\textcolor{blue}{#1}}

\newcommand{\bluefour}[1]{\textcolor{Blue4}{#1}}

\def\be{\begin{equation}}
\def\ee{\end{equation}}
\def\bea{\begin{eqnarray}}
\def\eea{\end{eqnarray}}

\begin{document}

\title{Kane-Lubensky phonons in Maxwell lattice frustrated Mott insulators}
\author{Hong-Hao Song}
\affiliation{International Center for Quantum Materials, School of Physics, 
Peking University, Beijing 100871, China}
\author{Gang v.~Chen}
\email{chenxray@pku.edu.cn}
\affiliation{International Center for Quantum Materials, School of Physics, 
Peking University, Beijing 100871, China}
\affiliation{Collaborative Innovation Center of Quantum Matter, 100871, Beijing, China}
\affiliation{Beijing Key Laboratory of Quantum Devices, Peking University, Beijing 100871, China}
\date{\today}


\begin{abstract}
We show that zero-energy gapless Weyl-line phonons
of the Kane-Lubensky's type can arise in the three-dimensional    
Maxwell lattice frustrated Mott insulators through the magnetopological mechanics.
In a pyrochlore antiferromagnet with the spin-lattice coupling, a magnetic field 
selects the spin state whose lattice distortion generates a $P4_3 32$ 
topological lattice. More crucially, the spin-lattice coupling and the spin 
configuration cause the bending of the neighbouring bonds, and converts the system 
into the topological Maxwell lattice. Remarkably, the resulting system 
is found to host the bulk zero-frequency Weyl-line phonons protected   
topologically, and these gapless phonons are not Goldstone modes. 
Unlike the conventional ${C_{\rm ph}\sim T^3}$ for the Goldstone phonons, 
these one-dimensional zero-mode manifolds yield a characteristic   
low-temperature phonon specific heat ${C_{\rm ph}\sim T^2}$. 
Our results could find applications in the Cr-based spinel systems,
and moreover, we establish a low-energy platform where an  
extensive number of topological zero-frequency phonons can 
strongly couple to other degrees of freedom, 
opening a route to exotic phonon-mediated phenomena. 
\end{abstract}

\maketitle

\noindent\bluefour{\it Introduction.}---Topological phononics extends the foundational concepts 
of topological band theory from electronic systems to lattice vibrations and mechanical waves, 
offering a robust principle for controlling phonon propagation with profound implications 
for thermal transport, interfacial dynamics, and phonon-based devices~\cite{https://doi.org/10.1002/adfm.201904784,Xue2022TopologicalAcoustics,Zhu_2023,https://doi.org/10.1002/adfm.202401684,chen2026topologicalphononics}. 
In crystalline solids, a rich variety of topological phonons have been theoretically 
predicted and experimentally observed. These are finite-frequency excitations  
with phonon band crossings or gap closures occurring at nonzero energies, 
where the topological character is encoded in Berry phases, 
Chern numbers, and other invariants associated with crystalline symmetries~\cite{PhysRevB.97.054305, PhysRevLett.120.016401, PhysRevLett.121.035302, PhysRevLett.123.245302,wang2022topological, PhysRevB.96.064106,jin2018recipe,PhysRevB.101.081403,PhysRevLett.126.185301,PhysRevLett.131.116602,doi:10.1126/science.adf8458}. 
They represent a direct phononic analogue of the topological phenomena long 
familiar in electronic systems.

Alongside this well-established analogue, 
a fundamentally different class of topological phonons  
has emerged from a distinct tradition,    
i.e. topological mechanics in Maxwell lattices~\cite{KaneLubensky2014,Lubensky_2015,PhysRevLett.117.068001,PhysRevLett.116.135503,Huber2016}. 
In a seminal work, Kane and Lubensky demonstrated that mechanical 
frames with a perfect balance between the number of degrees of freedom 
and the number of constraints, the so-called Maxwell condition, 
can possess nontrivial topological invariants that 
dictate the distribution of zero-frequency ``floppy modes" 
at boundaries~\cite{KaneLubensky2014}. 
Remarkably, this framework establishes a mapping 
from the bosonic phonon problem to a fermionic one 
by taking the square root of the dynamical matrix, 
revealing a topological classification that mirrors the
electronic topological insulators and semimetals~\cite{KaneLubensky2014, RevModPhys.82.3045, RevModPhys.90.015001}. 
These zero-frequency modes are {\sl not} Goldstone modes 
associated with the translational symmetry breaking. 
Instead, they are protected purely by lattice topology.  
To distinguish them from the conventional finite-energy 
topological phonons studied in natural crystals, 
we dub these ``square-root''-like phonons Kane-Lubensky phonons (KL phonons).

Despite the progress in topological mechanics, 
its materials' realization remains scarce. 
The pyrochlore antiferromagnets, 
long regarded as a fertile ground for frustrated magnetism~\cite{RevModPhys.82.53}, 
offer an intriguing yet unexplored arena. Decades of research have focused 
on their spin liquids, field-induced magnetization plateaux, 
and complex phase diagrams~\cite{PhysRevB.74.134409, PhysRevLett.88.067203,PhysRevB.66.064403,PhysRevLett.93.197203,PhysRevLett.80.2929,PhysRevLett.96.097207}. 
Early works established the  
${3:1}$ collinear plateau states, known as the $R$ state, 
in Cr-based spinels~\cite{PhysRevB.74.134409, PhysRevLett.96.097207,PhysRevLett.94.047202,Matsuda2007,PhysRevLett.104.047201}. 
These studies thoroughly 
characterized the magnetic order and spin-lattice coupling (SLC) effects~\cite{PhysRevLett.93.197203, PhysRevB.74.134409}, 
but did not recognize that the very lattice distortion produced by the SLC 
could transform the original pyrochlore network 
into a topological Maxwell lattice. 
This was primarily because their times 
were much ahead of the explosion of topological band physics. 
Indeed, the pyrochlore lattice is intrinsically a three-dimensional (3D) 
Maxwell lattice~\cite{PhysRevLett.117.068001}, and a suitable magnetic state, 
such as the field-induced $R$ state, 
can, through the imbalance of the bond energies, 
induce a static distortion that acquires a nontrivial topology. 
This magnetopological mechanics, 
recently illustrated in two-dimensional (2D) kagom\'e Mott insulators~\cite{song2026magnetopologicalmechanicsmaxwelllattice}, 
has not yet been applied to the pyrochlore family. 
The intricate interplay between the   
magnetism, the topological mechanics, and 
the lattice distortions 
remains overlooked, motivating the present work.

In this Letter, we fill the gap by demonstrating that the 
$R$-state induced lattice distortion in a pyrochlore antiferromagnet 
generates a $P4_3 32$ topological Maxwell lattice that hosts the 
KL phonons. In the KL phonon spectra, there exist the bulk zero-frequency 
Weyl-line phonon modes as well as the gapless edge modes. 
These topological excitations are not Goldstone modes; instead, 
they arise purely from lattice topology and form one-dimensional nodal lines 
in momentum space, with a linear dispersion in the transverse directions. 
We compute the momentum-space winding numbers and show that their jumps 
precisely trace the Weyl lines, which are further corroborated  
by the emergence of surface zero modes 
whose boundary localization shifts according to the topological index. 
Importantly, this Weyl-line phonon spectrum leads to a distinctive 
low-temperature specific heat ${C_{\rm ph}\sim T^2}$, 
providing a clear experimental signature that distinguishes 
it from the conventional Debye $T^3$-behaviour. 
Our results not only establish the pyrochlore antiferromagnets, 
particularly the Cr-based spinel systems, 
as a realistic 3D platform for topological KL phonons, 
but also open the door to strong low-energy electron-phonon couplings 
and emergent quantum phases driven by these soft topological modes.

\noindent\bluefour{\it SLC and bond bending.}---A perfect pyrochlore 
antiferromagnet has a $Fd\bar{3}m$ space group symmetry. 
Due to the balance of the forces from the neighbouring 180-degree bonds, 
the paramagnetic state does not have any topological mechanics. 
With a given magnetic order, the SLC 
distorts and bends the neighbouring bonds, converting a 
perfect pyrochlore lattice into a topological Maxwell lattice 
with the nontrivial KL phonons. This was dubbed magnetopological mechanics.   
To show this, we consider the following minimal model for the SLC, 
\begin{equation}
\mathcal{H} =
J \sum_{\langle ij\rangle}
\left(1+\gamma u_{ij}\right) 
{\bf S}_i\cdot{\bf S}_j 
- {\bf h}\cdot \sum_i  {\bf S}_i
+ \frac{k}{2}\sum_i |{\bf u}_i|^2 , 
\end{equation}
where $\langle ij \rangle$ denotes the nearest-neighbour bonds,   
and ${\bf S}_i$ is a local spin moment at the site $i$.   
The vector ${\bf u}_i$ denotes the displacement of the site $i$ 
from its equilibrium position ${\bf R}_i$, 
and ${R=|{\bf R}_i-{\bf R}_j|}$ is the equilibrium bond length. 
The parameter ${J>0}$ is the antiferromagnetic exchange 
in the undistorted lattice, and ${\gamma=(1/J)(dJ/dr)|_{r=R}}$ 
characterizes the linear dependence of the exchange coupling on the bond length. 
$u_{ij}=\hat{\bf e}_{ij}\cdot({\bf u}_i-{\bf u}_j)$ 
is the change of the bond length along the equilibrium bond direction, 
where $\hat{\bf e}_{ij}=({\bf R}_i-{\bf R}_j)/R$ is the bond unit vector. 
The second term is the Zeeman coupling to an external magnetic field ${\bf h}$, 
and $k$ is the elastic stiffness of the Einstein site phonons. 
This model describes the spin-only Mott insulator where
the spin interaction is primarily Heisenberg-like. 
For more generic moments and exchange interactions such as the 
strong spin-orbit-coupled Mott insulators,  
different forms of SLC should be considered.

The effective spin Hamiltonian is obtained 
by minimizing the Hamiltonian $\mathcal{H}$ 
with respect to the site displacements 
${\bf u}_i$~\cite{song2026magnetopologicalmechanicsmaxwelllattice,PhysRevB.74.134409, PhysRevB.105.174424}. 
This procedure yields the optimal site displacements
\begin{equation}
{\bf u}_i^\ast
=
-\frac{J\gamma}{k}
\sum_{j\in N(i)}
({\bf S}_i\cdot{\bf S}_j)\hat{\bf e}_{ji},
\label{eq:magnetopological_displacement}
\end{equation}
where $N(i)$ denotes the nearest neighbours of site $i$. 
This formula shows that the displacement of each site   
is determined by the imbalance of spin-bond energies around that site.
${\bf u}_i^\ast=0$ in the paramagnetic state, and could be nonzero 
in ordered states. 
A nonzero displacement bends the neighbouring bonds and establishes 
the connection between the magnetic states and the phonon spectra.


Substituting Eq.~\eqref{eq:magnetopological_displacement} 
into $\mathcal{H}$ leads to the effective spin Hamiltonian  
\begin{equation}
    \mathcal{H}_{\mathrm{eff}}
    =   \sum_{\langle ij \rangle} J
      \mathbf{S}_i \cdot \mathbf{S}_j 
      - \sum_i {\bf h}\cdot  {\bf S}_i
      -  J b \sum_{\langle ij \rangle}(\mathbf{S}_i \cdot \mathbf{S}_j)^2 
      + \mathcal{H}_{\mathrm{FN}},
    \label{eq:H_eff}
\end{equation}
where ${b = \frac{J \gamma^2}{k}}$ is a dimensionless SLC strength,
and the three-spin quartic term $\mathcal{H}_{\mathrm{FN}}$ takes  
the form
$   \mathcal{H}_{\mathrm{FN}}
    = - \frac{Jb}{2} 
      \sum_i \sum_{j \neq k \in \mathcal{N}(i)} 
      (\hat{\mathbf{e}}_{ij} \cdot \hat{\mathbf{e}}_{ik}) 
      (\mathbf{S}_i \cdot \mathbf{S}_j)
      (\mathbf{S}_i \cdot \mathbf{S}_k).
    \label{eq:H_FN}
$
The Zeeman fields and SLC-induced interactions in Eq.~\eqref{eq:H_eff} 
lift the extensive ground state degeneracy. 
In particular, the biquadratic spin exchange favours 
the collinear spin configurations 
and is responsible for the robust 1/2-magnetization plateau in
the intermediate field regime~\cite{PhysRevLett.93.197203,
PhysRevLett.94.047202,
PhysRevLett.96.097207,
PhysRevB.74.134409,
Matsuda2007,PhysRevLett.104.047201}.

\begin{figure}[b]
    \centering
    \begin{overpic}[width=\columnwidth]{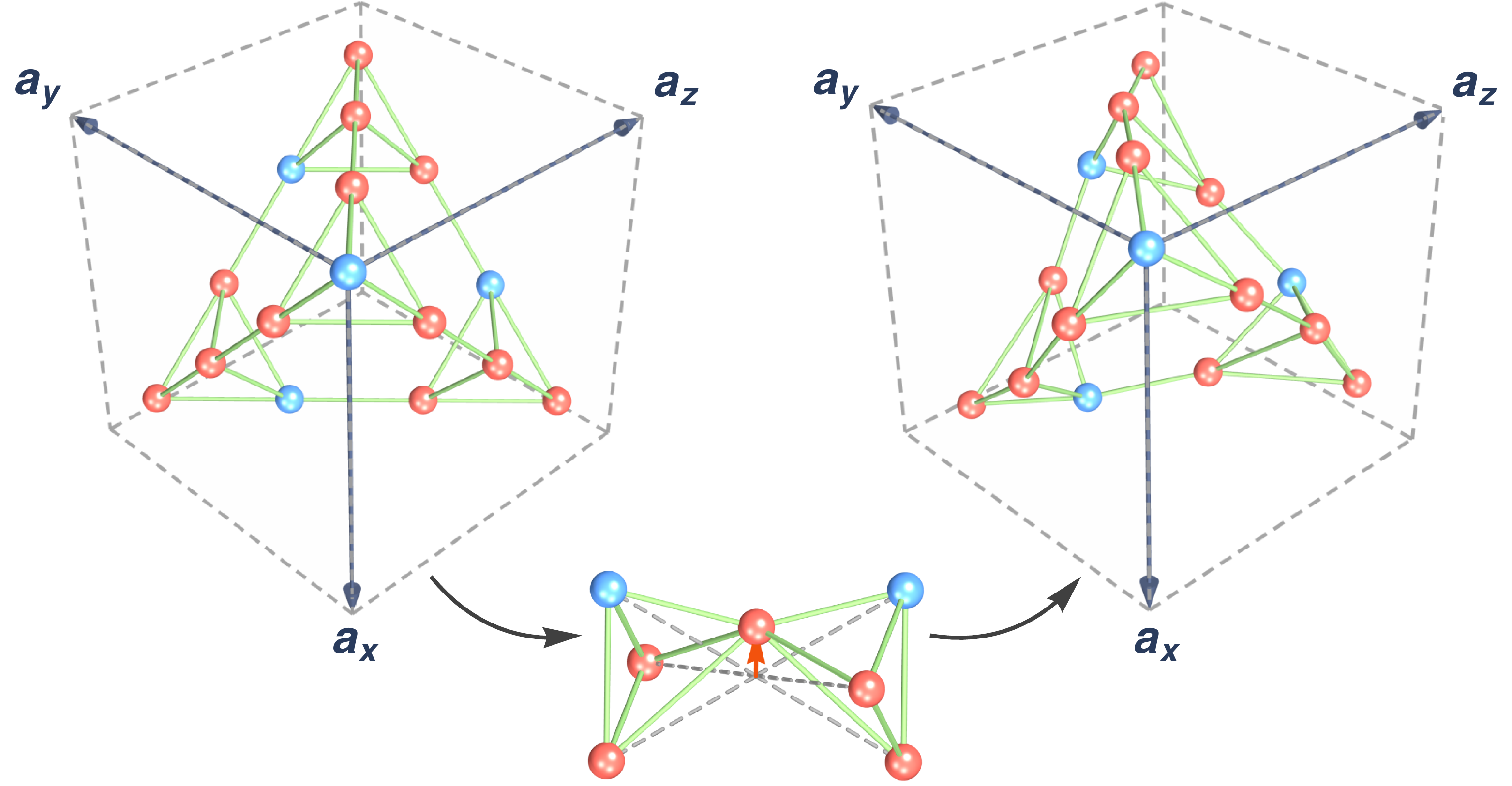}
        \put(1,50){\makebox(0,0)[lb]{(a)}}
        \put(51,50){\makebox(0,0)[lb]{(b)}}
    \end{overpic}

    \caption{
    $R$ state and the corresponding topological Maxwell lattice.
    (a) Cubic unit cell of the $R$-state spin configuration.
    (b) Topological lattice generated from the $R$-state distortion through SLC (inset).
    }
    \label{fig:Rstate_lattice}
\end{figure}

For the 1/2-magnetization plateau, each tetrahedron has the 3-up 1-down 
spin configurations that are referred as the ${3:1}$ collinear states and 
are favoured by the first three terms in $\mathcal{H}_{\mathrm{eff}}$. 
The extensive degeneracy of these states was lifted by the fourth term $\mathcal{H}_{\mathrm{FN}}$.
Restricted to the collinear spin configurations,  
$\mathcal{H}_{\mathrm{FN}}$ becomes simple Ising interactions between second
and third neighbor spins~\cite{SM, PhysRevB.74.134409}, 
and the favoured spin configuration for the 1/2-magnetization plateau is the $R$ state.
The $R$ state has a 16-site unit cell 
and is depicted in Fig.~\ref{fig:Rstate_lattice}(a). 
Based on Eq.~\eqref{eq:magnetopological_displacement}, 
the lattice distortion of the $R$ state is plotted in Fig.~\ref{fig:Rstate_lattice}(b),
and the distorted lattice has a $P4_332$ space group~\cite{SM, PhysRevB.74.134409,PhysRevLett.96.097207} 
and is found to be a topological Maxwell lattice below.


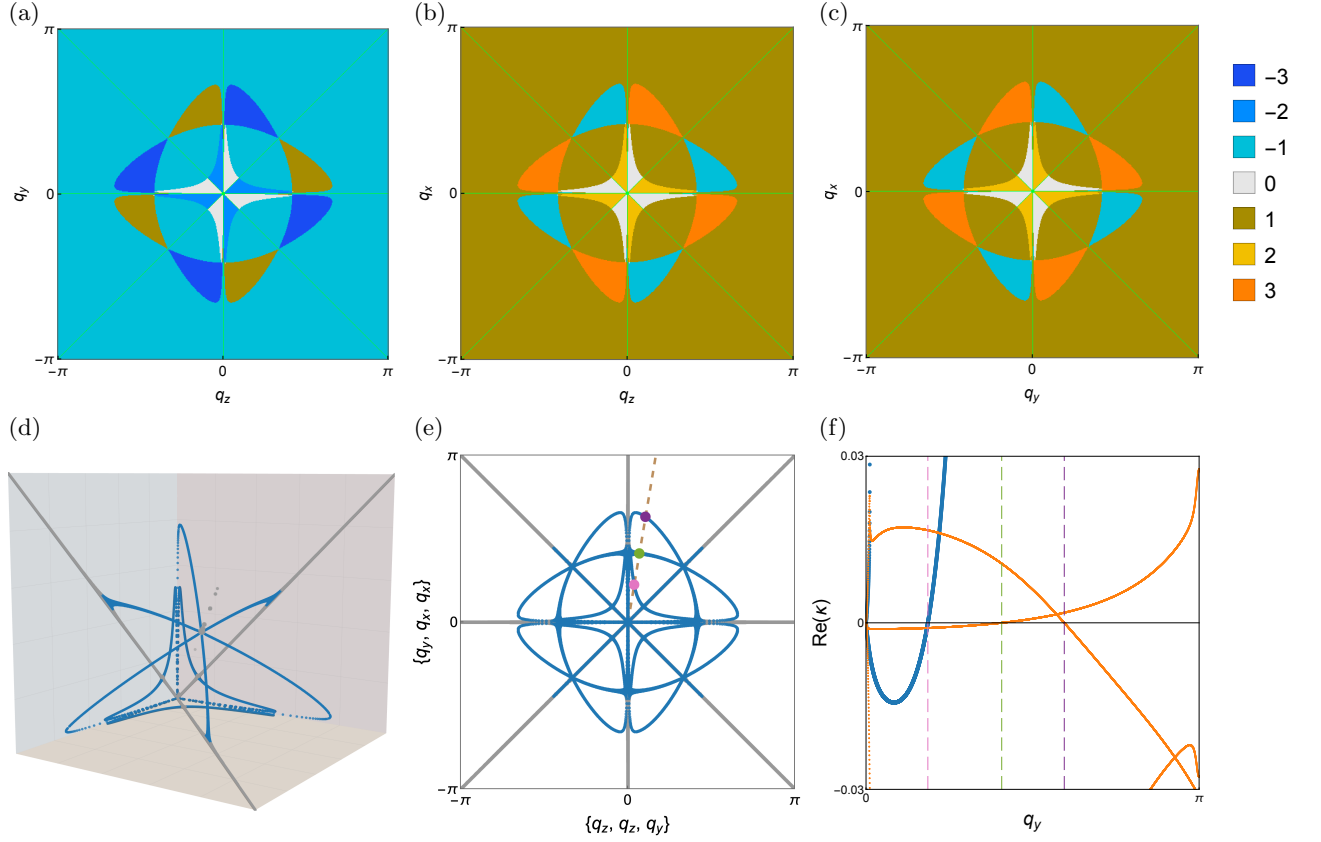
\begin{figure*}[!t]
    \centering

    \newcommand{\panelw}{0.285\textwidth}
    \newcommand{\legendw}{0.055\textwidth}
    \newcommand{\panelgap}{0.008\textwidth}

    \begin{minipage}[t]{\panelw}
        \centering
        \setlength{\unitlength}{0.01\linewidth}
        \begin{picture}(100,105)
            \put(0,-3){\includegraphics[width=\linewidth]{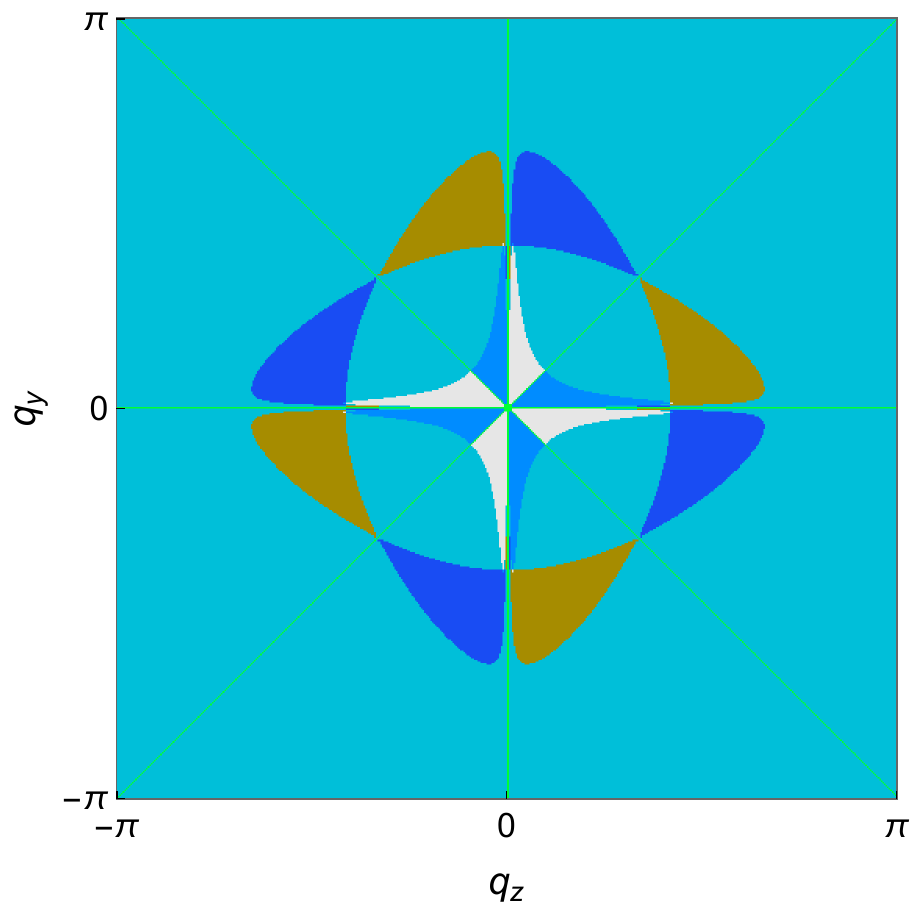}}
            \put(0,95){\makebox(0,0)[lb]{(a)}}
        \end{picture}
    \end{minipage}\hspace{\panelgap}
    \begin{minipage}[t]{\panelw}
        \centering
        \setlength{\unitlength}{0.01\linewidth}
        \begin{picture}(100,105)
            \put(0,-3){\includegraphics[width=\linewidth]{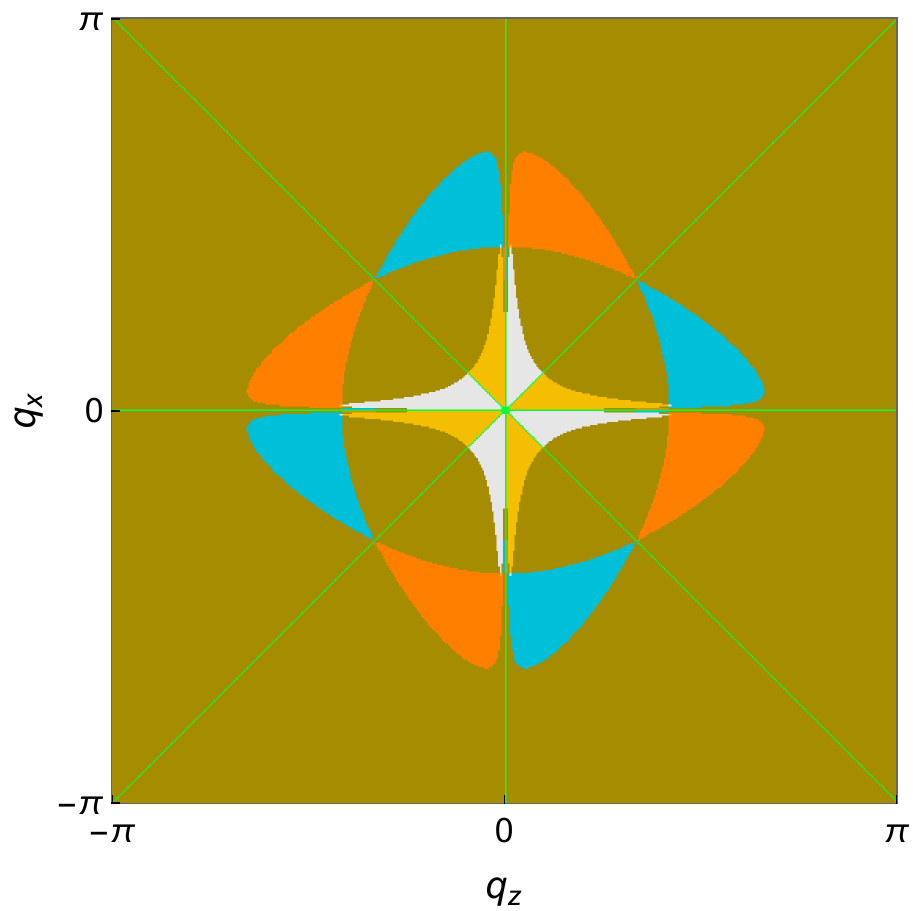}}
            \put(0,95){\makebox(0,0)[lb]{(b)}}
        \end{picture}
    \end{minipage}\hspace{\panelgap}
    \begin{minipage}[t]{\panelw}
        \centering
        \setlength{\unitlength}{0.01\linewidth}
        \begin{picture}(100,105)
            \put(0,-3){\includegraphics[width=\linewidth]{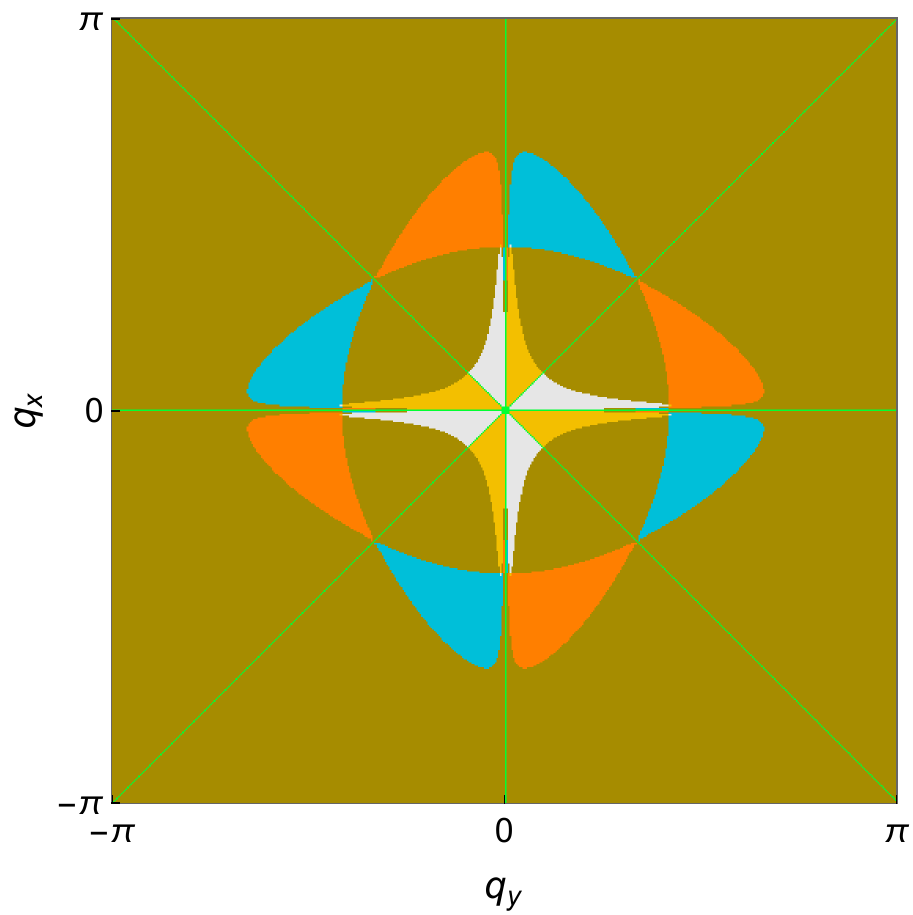}}
            \put(0,95){\makebox(0,0)[lb]{(c)}}
        \end{picture}
    \end{minipage}\hspace{\panelgap}
    \begin{minipage}[t]{\legendw}
        \centering
        \setlength{\unitlength}{0.01\linewidth}
        \begin{picture}(100,105)
            \put(0,107){\includegraphics[width=\linewidth]{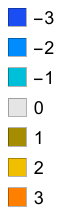}}
        \end{picture}
    \end{minipage}

    \vspace{0.25em}

    \begin{minipage}[t]{\panelw}
        \centering
        \setlength{\unitlength}{0.01\linewidth}
        \begin{picture}(100,105)
            \put(0,-1){\includegraphics[width=\linewidth]{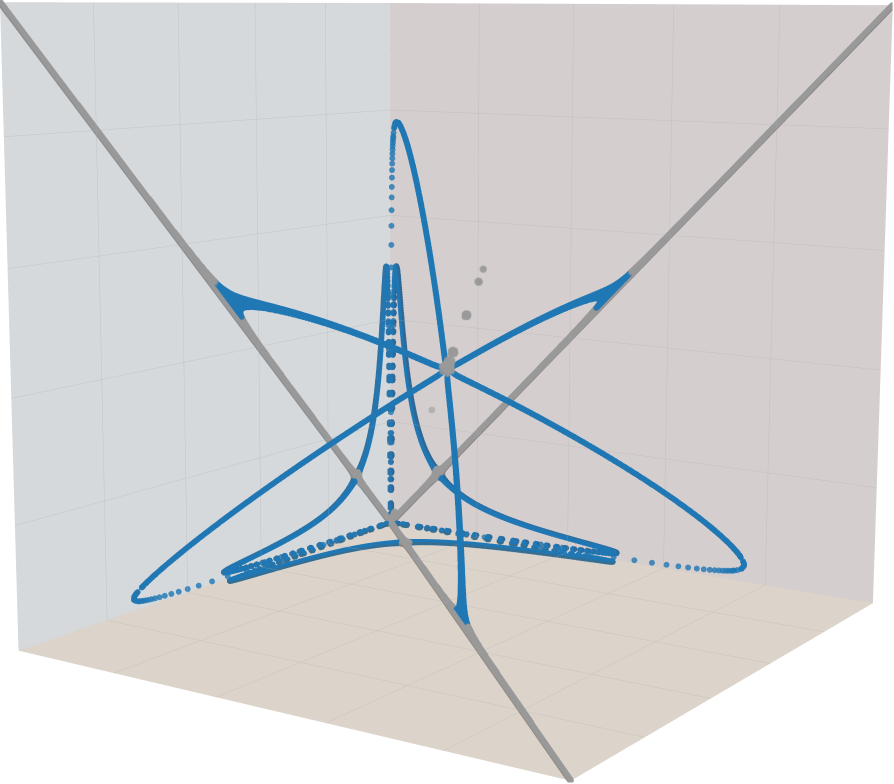}}
            \put(0,95){\makebox(0,0)[lb]{(d)}}
        \end{picture}
    \end{minipage}\hspace{\panelgap}
    \begin{minipage}[t]{\panelw}
        \centering
        \setlength{\unitlength}{0.01\linewidth}
        \begin{picture}(100,105)
            \put(-0.85,-7.6){\includegraphics[width=1.009\linewidth]{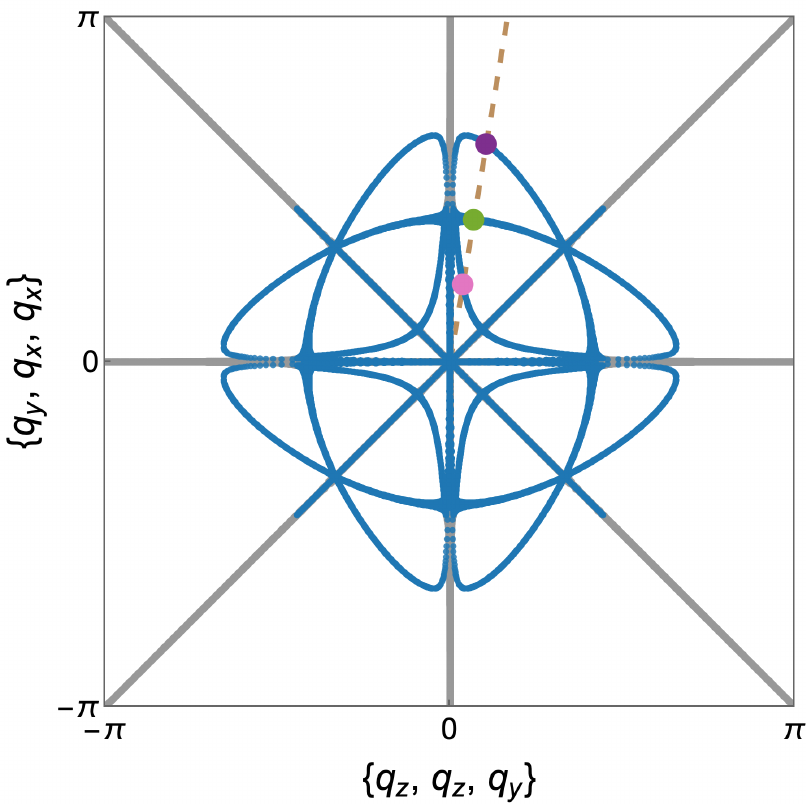}}
            \put(0,95){\makebox(0,0)[lb]{(e)}}
        \end{picture}
    \end{minipage}\hspace{\panelgap}
    \begin{minipage}[t]{\panelw}
        \centering
        \setlength{\unitlength}{0.01\linewidth}
        \begin{picture}(100,105)
            \put(-6.45,-9.45){\includegraphics[width=1.088\linewidth]{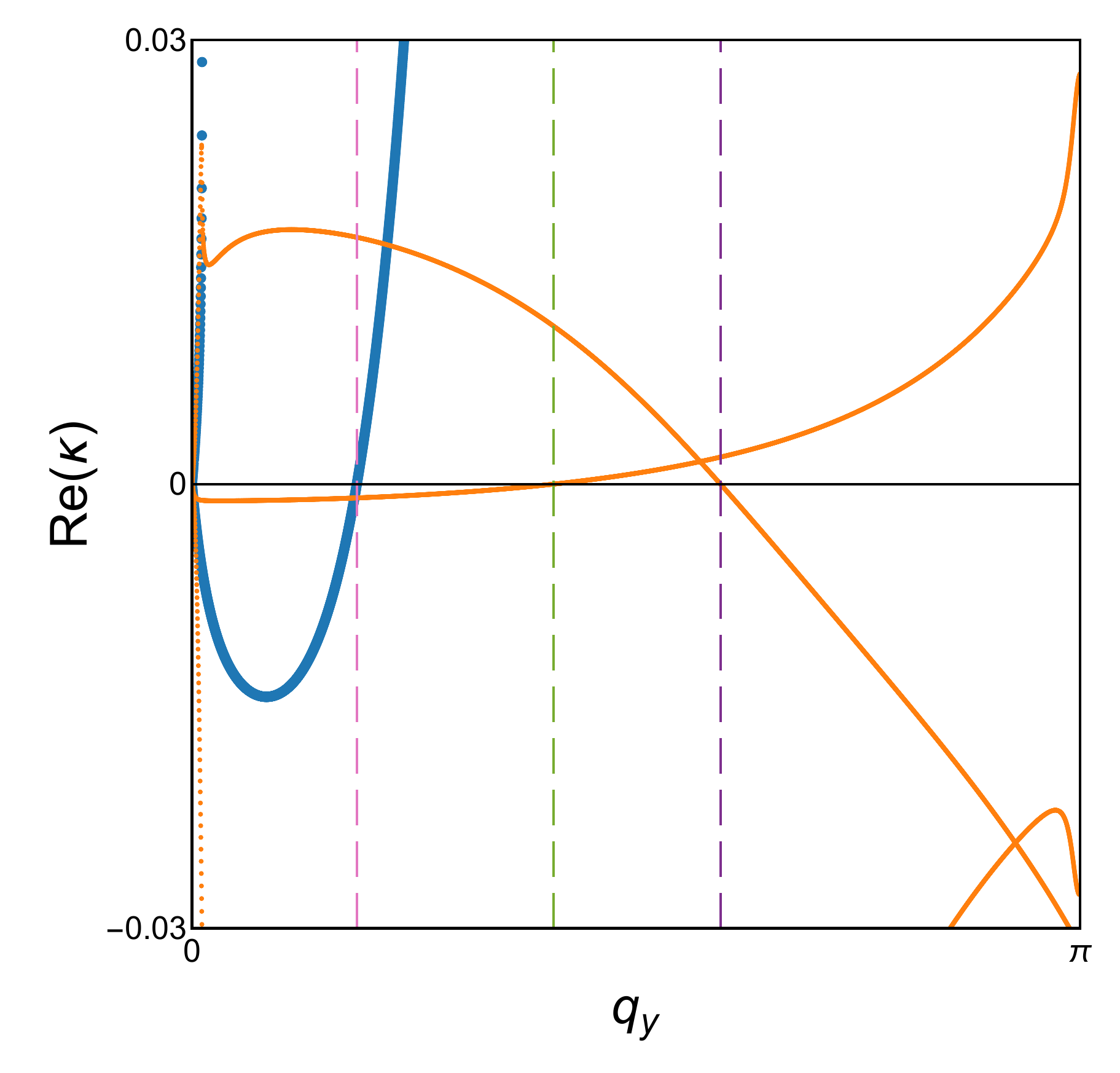}}
            \put(0,95){\makebox(0,0)[lb]{(f)}}
        \end{picture}
    \end{minipage}\hspace{\panelgap}
    \begin{minipage}[t]{\legendw}
        \mbox{}
    \end{minipage}
    \caption{
  Topological properties of the $R$-state induced topological Maxwell lattice.
    (a)--(c) Winding number diagrams in three momentum planes. 
    Colours denote different integer winding number values. 
    Boundaries between different winding number regions are projections of the Weyl lines, 
    whereas the green curves are projections of trivial zero-mode lines. 
    (d) Bulk zero-mode lines in the first octant of the Brillouin zone. 
    Blue (gray) curves denote topologically protected Weyl lines (trivial zero-mode lines). 
    (e) Projections of the bulk zero-mode lines in panel 
    (d) onto the planes perpendicular to ${\bf b}_x$, ${\bf b}_y$, and ${\bf b}_z$, 
    corresponding to the $q_y$-$q_z$, $q_x$-$q_z$, and $q_x$-$q_y$ planes, respectively. 
    The brown line marks the cut used in panel (f). 
    The coloured points mark the three intersections of the cut with the projected Weyl lines, 
    and the dashed lines of the same colors in panel (f) indicate the corresponding momenta.
    (f) Real parts of the complex inverse penetration depths along this cut. 
    Positive and negative values of $\mathrm{Re}(\kappa)$ correspond to the KL phonons 
    localized on the boundaries with outward normals $-{\bf a}_x$ and $+{\bf a}_x$, respectively. 
    The blue branch is non-degenerate, whereas the orange branches are twofold degenerate.
    }
    \label{fig:seven_panels}
\end{figure*}


\noindent\bluefour{\it Magnetopological mechanics.}---The SLC 
provides the underlying driving force for the lattice distortion.  
Once the new equilibrium positions are stabilized,  
one can perform the elasticity theory around the new equilibrium. 
Although the $R$ state helps distort the lattice and drive the topological lattice, 
the lattice dynamics and the spin dynamics are effectively independent at the quadratic level. 
At the lowest order of approximation, we rely on the topological lattice structure  
and directly compute the phonon spectrum~\cite{SM}.  
The equilibrium matrix $Q$ of the distorted lattice structure 
determines the phonon spectrum and the associated topological properties.
The topological properties of the phonons are associated with the winding numbers 
defined along the closed paths $C_i$ of $Q({\bf q})$ in the Brillouin zone, 
\begin{equation}
n_i = \frac{1}{2\pi i} \oint_{C_i} d\mathbf{q} \cdot \nabla_{\mathbf{q}} \ln \det Q(\mathbf{q}) .
\label{eq:ni_trace}
\end{equation}

The total number of zero-frequency KL phonon modes of the system 
is constrained by the Maxwell-Calladine index theorem~\cite{CALLADINE1978161,KaneLubensky2014}, 
${\nu = dim \ker Q^T - dim \ker Q}$, 
where $\ker Q$ and $\ker Q^T$ are the kernels of $Q$ and its transpose, 
respectively. 
Here, \(\nu\) is the difference between the numbers of displacement 
degrees of freedom and bond constraints, 
while \(\ker Q^T\) and \(\ker Q\) represent the zero-frequency modes 
and states of self-stress, respectively. 
For a periodic Maxwell lattice, \(\nu=0\),  
thus the number of zero modes equals the number of states of self-stress.
A lattice distortion can remove the states of self-stress, 
and gap out their associated bulk zero modes.
In 2D topological Maxwell lattices for the kagom\'e geometry, 
the bulk phonon spectrum is found to be ``fully gapped'' except 
for the Goldstone zero modes at ${\bf q}=0$~\cite{song2026magnetopologicalmechanicsmaxwelllattice},
and the bulk topology is characterized by winding numbers 
Eq.~\eqref{eq:ni_trace}. 
A finite sample could therefore host zero-frequency KL phonons 
localized at its boundaries.

The $R$-state induced 3D topological Maxwell lattice 
has a much richer structure. 
This lattice distortion does not completely
gap out all the non-Goldstone zero modes.
The remaining zero modes generically form Weyl lines in 3D Maxwell lattices~\cite{SM}.
For each primitive reciprocal lattice vector ${\bf b}_i$, 
we evaluate the winding number in Eq.~\eqref{eq:ni_trace} 
along the path $C_i$ from ${\bf q}_{\perp}$ to ${\bf q}_{\perp}+{\bf b}_i$, 
where ${\bf q}_{\perp}$ lies in the plane 
through the origin perpendicular to ${\bf b}_i$. 
For a fully gapped topological Maxwell lattice, 
the winding number $n_i$ is independent of ${\bf q}_{\perp}$. 
Here, it, however, depends on ${\bf q}_{\perp}$, 
owing to the Weyl-line structure of the remaining zero modes. 



In Figs.~\ref{fig:seven_panels}(a)--(c), we depict the winding numbers 
that are obtained by scanning ${\bf q}_{\perp}$ over the planes 
perpendicular to the three primitive reciprocal vectors. 
These integer-valued winding numbers are piecewise constant, 
and the boundaries between different regions are the projections of the Weyl lines. 
Across each boundary, the change in the winding number is
equal to the net topological charge of the crossed Weyl lines, 
as in the nodal-line semimetals~\cite{PhysRevLett.117.068001,PhysRevB.84.235126,PhysRevLett.115.036806}. 
The Weyl-line structure is therefore stable 
and can disappear only through annihilation 
with a Weyl line of opposite charge.



Owing to the remaining cubic symmetry of the \(P4_3 32\) space group~\cite{SM}, 
the winding number diagrams in different momentum planes  
of Figs.~\ref{fig:seven_panels}(a)-(c) share the same  structure~\cite{MaradudinVosko1968,DresselhausGroupTheory2008}.
Despite this property, the winding numbers along the $x$ direction 
differ from those along the $y$ and $z$ directions  
based on the unit-cell dipole ${\bf R}_L=(-1,1,1)$~\cite{SM}. 
This dipole only shifts their overall integer values, 
leaving the winding number jumps unchanged. 
To locate the Weyl lines in momentum space, 
we scan the Brillouin zone for the zero-frequency phonons. 
Due to the remaining cubic symmetry, it is sufficient to 
consider only the first octant. The resulting zero modes 
form the one-dimensional lines that are shown in Fig.~\ref{fig:seven_panels}(d). 
Their projections along ${\bf b}_x$, ${\bf b}_y$, and ${\bf b}_z$ 
are displayed in Fig.~\ref{fig:seven_panels}(e). 
The blue lines project onto the boundaries 
between distinct winding number regions 
in Figs.~\ref{fig:seven_panels}(a)--(c), 
and are therefore topologically protected Weyl lines. 
In contrast, the gray lines project onto the green lines, 
across which the winding number stays unchanged, 
and are topologically trivial.

Finally, we analyze the boundary KL phonons. 
For concreteness, we consider a boundary whose normal is parallel to the primitive lattice vector ${\bf a}_x$. 
At the fixed boundary momentum ${\bf q}_{\parallel}=q_y{\bf b}_y+q_z{\bf b}_z$, 
the complex inverse penetration depths $\kappa$ are obtained by solving
$\det Q^{\dagger}(i\kappa{\bf b}_x+{\bf q}_{\parallel})=0$~\cite{SM, doi:10.1137/S0036144500381988}. 
Each solution describes a boundary zero mode that varies 
along the $x$ direction as $e^{-\kappa x}$, 
and $|\mathrm{Re}(\kappa)|^{-1}$ gives its penetration depth~\cite{Lubensky_2015,PhysRevLett.117.068001}. 
With this convention, $\mathrm{Re}(\kappa)>0$ and $\mathrm{Re}(\kappa)<0$ 
correspond to the phonon modes that are localized on the boundaries   
with the outward normals $-{\bf a}_x$ and $+{\bf a}_x$, respectively. 
Fig.~\ref{fig:seven_panels}(f) shows $\mathrm{Re}(\kappa)$ 
along the brown cut in the $q_y$-$q_z$ plane of Fig.~\ref{fig:seven_panels}(e). 
The blue dotted line corresponds to the non-degenerate solution, 
whereas the orange dotted line is twofold degenerate. 
For each fixed ${\bf q}_{\parallel}$, there are eight solutions in total, 
corresponding to the eight boundary modes 
required by the Maxwell-Calladine index theorem~\cite{SM}. 
Their distribution between the two boundaries is determined by the boundary index theorem 
${\nu_B=\nu_L+\nu_T}$~\cite{KaneLubensky2014}. 
Here $\nu_L$ depends on the boundary termination, 
whereas $\nu_T$ is determined by the bulk winding numbers
$\nu_T =
\frac{{\bf G}}{2\pi}\cdot
\sum_i n_i {\bf a}_i $. 
For the two opposite boundaries, 
${\bf G}=\pm{\bf b}_x$ gives the opposite values of $\nu_T$. 
A winding number jump therefore increases the zero-mode count on one boundary 
and decreases it by the same amount on the other, leaving the total count unchanged. 
Along the brown cut, the winding numbers in Fig.~\ref{fig:seven_panels}(a) give
$
\nu_T: 0\rightarrow -1\rightarrow -3\rightarrow -1
$
for the boundary with ${\bf G}={\bf b}_x$. 
Correspondingly, Fig.~\ref{fig:seven_panels}(f) 
shows that the numbers of solutions with positive and negative $\mathrm{Re}(\kappa)$ 
change as
$
3/5\rightarrow 4/4\rightarrow 6/2\rightarrow 4/4 ,
$
revealing the transfer of zero modes between the two boundaries~\cite{SM}. 
This provides a boundary manifestation of the bulk Weyl-line topology. 
These boundary KL phonons are analogous to the drumhead surface 
states of the electron nodal-line semimetals, where projected nodal loops similarly 
separate regions with different one-dimensional winding numbers~\cite{PhysRevB.84.235126}. 
However, unlike generic drumhead states, whose exact flatness is not protected, 
the topological Maxwell structure pins the KL phonons to zero frequency.

\noindent\bluefour{\it Discussions.}---In this Letter, we have realized the zero-frequency KL phonons 
in the bulk of 3D Maxwell lattice frustrated Mott insulators through 
the magnetopological mechanics, whereas the previous 2D realizations only 
supported boundary zero-frequency modes. 
These stable bulk KL phonons form the topologically protected Weyl line structures 
and yield a characteristic low-temperature phonon specific heat 
$C_{\rm ph}\sim T^2$, providing a direct experimental signature.
Experimentally, the required pyrochlore-based topological Maxwell lattice can 
be accessed on the field-induced ${1/2}$-magnetization plateau, 
and the SLC selects the $R$ state and induces the associated lattice distortion. 
The Cr-based spinels exhibiting this plateau are therefore promising candidate systems.
From the materials' perspective, 
the effects of the SLC and other microscopic interactions 
can be incorporated into the topological lattice model through renormalized bond stiffnesses. 
As long as the resulting stiffness matrix remains positive definite, 
the zero modes and winding numbers remain unchanged~\cite{SM}.

The realization of a large number of non-Goldstone zero-frequency and/or low-energy phonons 
in realistic materials opens new directions for the correlated electronic physics. 
In the metallic realizations, the coupling between the itinerant electrons and the Weyl-line KL phonons 
could strongly enhance the low-energy electron-phonon interactions and promote
the superconductivity, potentially raising the transition temperature through the enhanced 
low-energy phonon density of states. 
The same coupling may also strongly scatter the Landau quasiparticles and produce non-Fermi-liquid behavior. 
More broadly, the magnetopological mechanics provides a tunable platform in which magnetic order, 
lattice topology, and the electronic degrees of freedom intertwine, offering a 
route to quantum phases driven by the topological zero-frequency phonons.

\noindent\emph{Acknowledgments.}---This work is supported by BJNSF with No. F261004, 
and NSFC with Grants No.~92565110 and No.~12574061. 

\bibliographystyle{apsrev4-2}
\bibliography{Refs}

\clearpage
\newpage
\widetext

\begin{center}
\textbf{\large Supplemental Materials for \\ ``Kane-Lubensky phonons in Maxwell lattice frustrated Mott insulators''}
\end{center}

\addtocontents{toc}{\protect\setcounter{tocdepth}{0}}
{
\tableofcontents
}

\renewcommand{\thefigure}{S\arabic{figure}}
\renewcommand{\theHfigure}{S\arabic{figure}}
\setcounter{figure}{0}
\renewcommand{\theequation}{S\arabic{equation}}
\renewcommand{\theHequation}{S\arabic{equation}}
\setcounter{equation}{0}
\renewcommand{\thesection}{\Roman{section}}
\setcounter{section}{0}
\setcounter{secnumdepth}{4}

\section{The $R$ State and the Induced Topological Lattice}
\label{app:Rstate}
\label{app:Rstate_geometry}

\subsection{The $R$ State}
\label{app:Rstate_configuration}
\begin{figure}[!htbp]
    \centering
    \begin{minipage}{0.68\columnwidth}
    \centering

    \begin{minipage}[t]{0.38\linewidth}
        \centering
        \setlength{\unitlength}{0.01\linewidth}
        \begin{picture}(80,105)
            \put(0,0){\includegraphics[width=\linewidth]{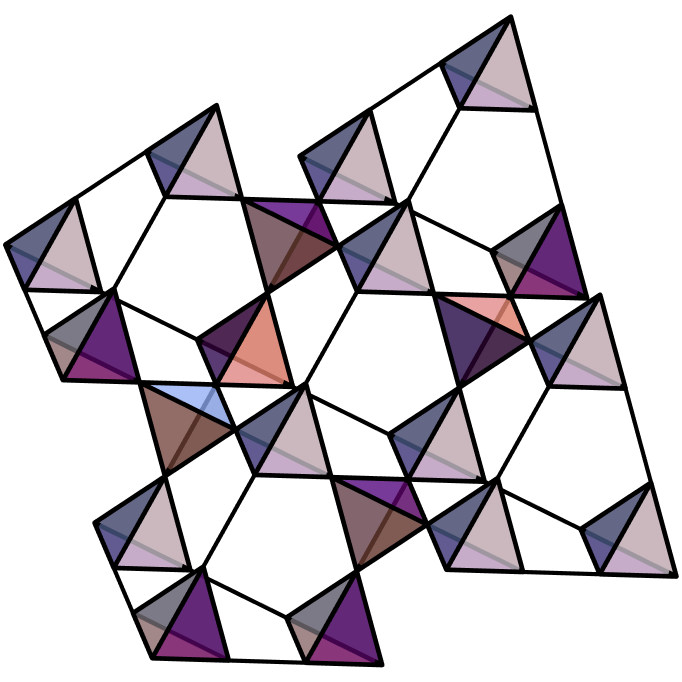}}
            \put(0,95){\makebox(0,0)[lb]{(a)}}
        \end{picture}
    \end{minipage}
    \hfill
    \begin{minipage}[t]{0.38\linewidth}
        \centering
        \setlength{\unitlength}{0.01\linewidth}
        \begin{picture}(120,105)
            \put(-10,15){\includegraphics[width=\linewidth]{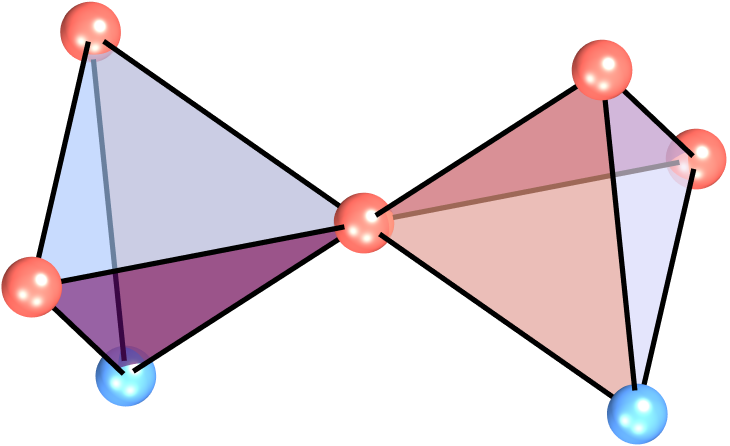}}
            \put(-10,95){\makebox(0,0)[lb]{(b)}}
        \end{picture}
    \end{minipage}

    \vspace{0.5em}

    \begin{minipage}[t]{0.38\linewidth}
        \centering
        \setlength{\unitlength}{0.01\linewidth}
        \begin{picture}(80,105)
            \put(0,15){\includegraphics[width=\linewidth]{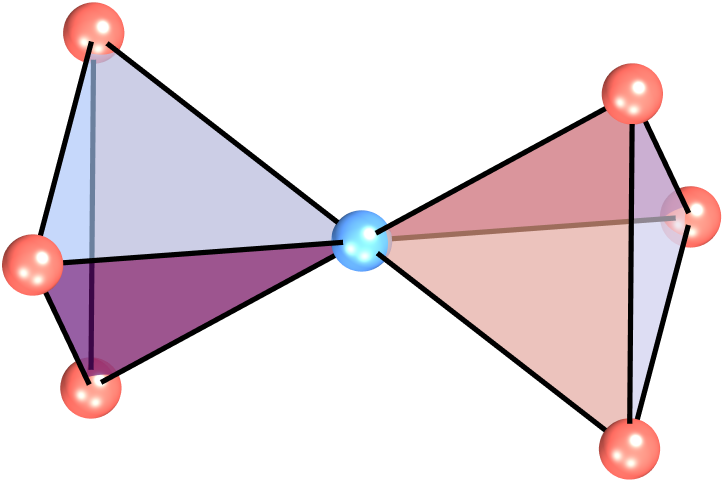}}
            \put(0,95){\makebox(0,0)[lb]{(c)}}
        \end{picture}
    \end{minipage}
    \hfill
    \begin{minipage}[t]{0.38\linewidth}
        \centering
        \setlength{\unitlength}{0.01\linewidth}
        \begin{picture}(120,105)
            \put(-10,15){\includegraphics[width=\linewidth]{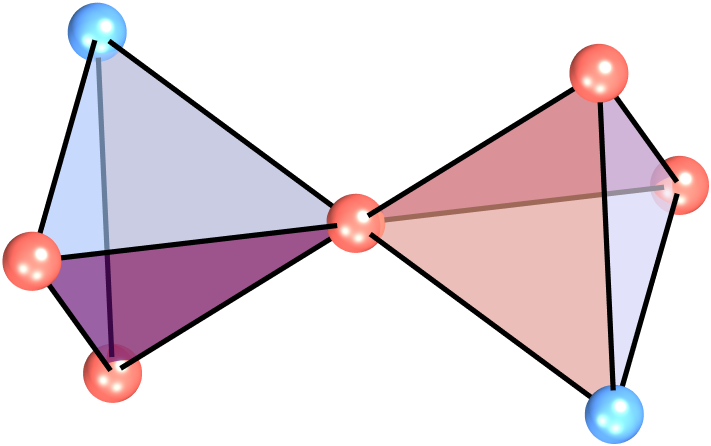}}
            \put(-10,95){\makebox(0,0)[lb]{(d)}}
        \end{picture}
    \end{minipage}
    \end{minipage}

    \caption{
    Pyrochlore lattice and local spin configurations in the $3\!:\!1$ manifold.
    (a) The pyrochlore lattice.
    (b) Local configuration around a spin-up site that induces a nonzero displacement ${\bf u}_i^\ast$.
    (c) Local configuration around a spin-down site.
    (d) Local configuration around a spin-up site with ${\bf u}_i^\ast=0$.
    Red and blue spheres denote spin-up and spin-down sites, respectively.
    }
    \label{fig:app_lattice_pattern}
\end{figure}

For the pyrochlore lattice shown in
Fig.~\ref{fig:app_lattice_pattern}(a), the effective spin Hamiltonian in Eq.~\eqref{eq:H_eff}
can be written as
\begin{equation}
    \mathcal{H}_{\mathrm{eff}}
    = \frac{J}{2}\sum_t
    \left[
        \left(\mathbf S_t-\frac{\mathbf h}{2J}\right)^2
        -\left(\frac{\mathbf h}{2J}\right)^2
    \right]
    -Jb\sum_{\langle i,j\rangle}
    \left(\mathbf S_i\cdot\mathbf S_j\right)^2
    +\mathcal H_{\mathrm{FN}}.
    \label{eq:app_H_eff}
\end{equation}
Here $\mathbf S_t=\sum_{j\in t}\mathbf S_j$ is the total spin of
tetrahedron $t$, $\mathbf h=h\hat{\mathbf z}$,  and
$\mathcal H_{\mathrm{FN}}$ denotes the three-spin quartic interaction discussed in the main text.
The first term in Eq.~\eqref{eq:app_H_eff} is minimized when
$\mathbf S_t=\mathbf h/(2J)$ on every tetrahedron.
This constraint, together with the collinearity favored by the biquadratic terms, stabilizes the 3-up 1-down arrangement on each tetrahedron around $h=4J$, giving rise to the collinear $3:1$ manifold.

Around each site, the possible spin configurations are shown in Fig.~\ref{fig:app_lattice_pattern}(b)--(d), where the red and blue spheres represent spin-up and spin-down respectively.
The $3:1$ manifold can be constructed by randomly assembling these three types of local configurations, which gives a large degeneracy. 
The ground state can be determined by energetically selecting from the \(3:1\) manifold. To make this selection explicit, we rewrite the effective Hamiltonian as~\cite{PhysRevB.74.134409}
\begin{equation}
\mathcal{H}_{\rm eff}
=
J\sum_{\langle ij\rangle}\sigma_i\sigma_j
- h\sum_i\sigma_i
-
\frac{k}{2}\sum_i \left|{\bf u}_i^{*}\right|^2 .
\end{equation}
Here $\sigma_i$ is an Ising variable: ${\bf S}_i=\sigma_i\hat{\bf z}$, with
$\sigma_i=+1$ ($-1$) for an up (down) spin.
Within the \(3:1\) manifold, the first two terms are constant, so the energy is lowered by maximizing the distortion. For a spin-up site (red site), the surrounding spin configuration in Fig.~\ref{fig:app_lattice_pattern}(b) induces a nonzero displacement ${\bf u}_i^{*}$, whereas Fig.~\ref{fig:app_lattice_pattern}(d) gives ${\bf u}_i^{*}=0$. For a spin-down site (blue site), the surrounding spin configuration is unique, as shown in Fig.~\ref{fig:app_lattice_pattern}(c). The ground state is therefore constructed from globally compatible type-(b) and type-(c) local configurations.
This is the so-called $R$ state.

Figure~\ref{fig:Rstate_lattice}(a) in the main text shows the cubic unit cell of the $R$ state, which contains 16 sites.
Each lattice site is labeled by $(\mathbf R,\alpha)$, where $\mathbf R=(n_x,n_y,n_z)$ denotes the unit cell and $\alpha=1,\ldots,16$ labels the site in a unit cell.
The position of each site $(\mathbf R,\alpha)$ is
\begin{equation}
\mathbf r_{\mathbf R\alpha}^{(0)}
=
\mathbf R+\mathbf r_\alpha^{(0)} .
\end{equation}
The first three columns of Table~\ref{tab:Rstate_site_distortions} give the index, spin direction, and position of every site.

The cubic unit cell contains 48 nearest-neighbor bonds: 30 intracell bonds and 18 bonds connecting neighboring cells.
To construct the equilibrium matrix $Q$, we assign an orientation to each bond.
For a bond oriented from $(\mathbf R_1,\alpha_1)$ to $(\mathbf R_2,\alpha_2)$, the corresponding bond vector is
\begin{equation}
\mathbf d_{12}^{(0)}
=
\mathbf R_2+\mathbf r_{\alpha_2}^{(0)}
-\mathbf R_1-\mathbf r_{\alpha_1}^{(0)} .
\label{eq:app_undistorted_bond}
\end{equation}
The first four columns of Table~\ref{tab:bond_configuration} list the bond IDs, endpoint pairs,
and bond vectors for all 48 bonds.

\subsection{The $R$-State Induced Topological Lattice}
\label{app:Rstate_topological_lattice}

According to magnetopological mechanics, the spin-lattice coupling (SLC) can generate a topological lattice. For each site \(i\), the SLC-induced displacement is given by
\begin{equation}
{\bf u}_i^\ast
=
-\frac{J\gamma}{k}
\sum_{j\in N(i)}
({\bf S}_i\cdot{\bf S}_j)\hat{\bf e}_{ji}.
\label{SMeq:magnetopological_displacement}
\end{equation}
The SLC-induced topological lattice is parameterized as
\begin{equation}
\mathbf r_{\mathbf R\alpha}
=
\mathbf R+\mathbf r_\alpha^{(0)}
+2\frac{J\abs{\gamma}}{k} \boldsymbol{\eta}_\alpha ,
\label{eq:app_site_distortion}
\end{equation}
where $\boldsymbol{\eta}_\alpha$ is the dimensionless distortion direction fixed by the $R$-state.
The calculations in the main text use
$2\frac{J\abs{\gamma}}{k}=0.08$.
The last two columns of Table~\ref{tab:Rstate_site_distortions} give the distortion direction and distorted position of each site.
\begin{table*}[t]
\centering
\caption{Site data for the $R$-state induced topological lattice at
$2 \frac{J\abs{\gamma}}{k}=0.08$, with
$\Delta=2\frac{J\abs{\gamma}}{\sqrt{2}k}=\frac{\sqrt{2}}{25}$ and
$\mathbf r_\alpha=\mathbf r_\alpha^{(0)}+2\frac{J\abs{\gamma}}{k}\boldsymbol{\eta}_\alpha$.}
\label{tab:Rstate_site_distortions}
\label{tab:Rstate}
\small
\setlength{\tabcolsep}{5.5pt}
\renewcommand{\arraystretch}{1.18}
\begin{ruledtabular}
\begin{tabular}{c c c @{\hspace{0.8em}} c c}
$\alpha$ & Spin & $\mathbf r_\alpha^{(0)}$ &
$\boldsymbol{\eta}_\alpha$ & $\mathbf r_\alpha$ \\
\hline
1  & $\downarrow$ & $(0,0,0)$
   & $(0,0,0)$ & $(0,0,0)$ \\
2  & $\uparrow$ & $(0,\frac14,\frac14)$
   & $(-\frac{1}{\sqrt2},0,-\frac{1}{\sqrt2})$
   & $(-\Delta,\frac14,\frac14-\Delta)$ \\
3  & $\uparrow$ & $(\frac14,0,\frac14)$
   & $(-\frac{1}{\sqrt2},-\frac{1}{\sqrt2},0)$
   & $(\frac14-\Delta,-\Delta,\frac14)$ \\
4  & $\uparrow$ & $(\frac14,\frac14,0)$
   & $(0,-\frac{1}{\sqrt2},-\frac{1}{\sqrt2})$
   & $(\frac14,\frac14-\Delta,-\Delta)$ \\
5  & $\uparrow$ & $(0,\frac12,\frac12)$
   & $(0,\frac{1}{\sqrt2},-\frac{1}{\sqrt2})$
   & $(0,\frac12+\Delta,\frac12-\Delta)$ \\
6  & $\uparrow$ & $(0,\frac34,\frac34)$
   & $(\frac{1}{\sqrt2},\frac{1}{\sqrt2},0)$
   & $(\Delta,\frac34+\Delta,\frac34)$ \\
7  & $\uparrow$ & $(\frac14,\frac12,\frac34)$
   & $(\frac{1}{\sqrt2},0,-\frac{1}{\sqrt2})$
   & $(\frac14+\Delta,\frac12,\frac34-\Delta)$ \\
8  & $\downarrow$ & $(\frac14,\frac34,\frac12)$
   & $(0,0,0)$
   & $(\frac14,\frac34,\frac12)$ \\
9  & $\uparrow$ & $(\frac12,0,\frac12)$
   & $(-\frac{1}{\sqrt2},0,\frac{1}{\sqrt2})$
   & $(\frac12-\Delta,0,\frac12+\Delta)$ \\
10 & $\downarrow$ & $(\frac12,\frac14,\frac34)$
   & $(0,0,0)$
   & $(\frac12,\frac14,\frac34)$ \\
11 & $\uparrow$ & $(\frac34,0,\frac34)$
   & $(0,\frac{1}{\sqrt2},\frac{1}{\sqrt2})$
   & $(\frac34,\Delta,\frac34+\Delta)$ \\
12 & $\uparrow$ & $(\frac34,\frac14,\frac12)$
   & $(-\frac{1}{\sqrt2},\frac{1}{\sqrt2},0)$
   & $(\frac34-\Delta,\frac14+\Delta,\frac12)$ \\
13 & $\uparrow$ & $(\frac12,\frac12,0)$
   & $(\frac{1}{\sqrt2},-\frac{1}{\sqrt2},0)$
   & $(\frac12+\Delta,\frac12-\Delta,0)$ \\
14 & $\uparrow$ & $(\frac12,\frac34,\frac14)$
   & $(0,-\frac{1}{\sqrt2},\frac{1}{\sqrt2})$
   & $(\frac12,\frac34-\Delta,\frac14+\Delta)$ \\
15 & $\downarrow$ & $(\frac34,\frac12,\frac14)$
   & $(0,0,0)$
   & $(\frac34,\frac12,\frac14)$ \\
16 & $\uparrow$ & $(\frac34,\frac34,0)$
   & $(\frac{1}{\sqrt2},0,\frac{1}{\sqrt2})$
   & $(\frac34+\Delta,\frac34,\Delta)$ \\
\end{tabular}
\end{ruledtabular}
\end{table*}

For the same bond orientation introduced above, the distorted bond vector is
\begin{equation}
\mathbf d_{12}
=
\mathbf d_{12}^{(0)}
+2\frac{J \abs{\gamma}}{k}\left(\boldsymbol{\eta}_{\alpha_2}-\boldsymbol{\eta}_{\alpha_1}\right).
\label{eq:app_distorted_bond}
\end{equation}
The unit vector is
$
\widehat{\mathbf d}_{12}
=
\frac{\mathbf d_{12}}{|\mathbf d_{12}|}$.
The final vector column in Table~\ref{tab:bond_configuration} gives the
distorted bond vector \(\mathbf d_{12}\) for
\(2\frac{J\abs{\gamma}}{k}=0.08\). 
Tables~\ref{tab:Rstate_site_distortions} and
\ref{tab:bond_configuration} completely specify the periodic
topological lattice used in the phonon calculation.

\begin{table*}[t]
\centering
\caption{The 48 directed bonds of the cubic unit cell. The two
endpoints and the undistorted and distorted bond vectors are listed at
$2\frac{J \abs{\gamma}}{k}=0.08$, with $\Delta=\sqrt{2}/25$.}
\label{tab:bond_configuration}
\label{tab:Rstate_bond_vectors}
\scriptsize
\setlength{\tabcolsep}{2.2pt}
\renewcommand{\arraystretch}{1.22}
\resizebox{\textwidth}{!}{%
\begin{tabular}{c c c c c @{\hspace{1em}} c c c c c}
\hline\hline
Bond & $(\mathbf R_1,\alpha_1)$ & $(\mathbf R_2,\alpha_2)$ &
$\mathbf d_{12}^{(0)}$ & $\mathbf d_{12}$ &
Bond & $(\mathbf R_1,\alpha_1)$ & $(\mathbf R_2,\alpha_2)$ &
$\mathbf d_{12}^{(0)}$ & $\mathbf d_{12}$ \\
\hline
1 &(0,0,0;1)&(0,0,0;2)&$(0,\frac14,\frac14)$&$(-\Delta,\frac14,\frac14-\Delta)$&25&(0,0,0;13)&(0,0,0;14)&$(0,\frac14,\frac14)$&$(-\Delta,\frac14,\frac14+\Delta)$\\
2 &(0,0,0;1)&(0,0,0;3)&$(\frac14,0,\frac14)$&$(\frac14-\Delta,-\Delta,\frac14)$&26&(0,0,0;13)&(0,0,0;15)&$(\frac14,0,\frac14)$&$(\frac14-\Delta,\Delta,\frac14)$\\
3 &(0,0,0;1)&(0,0,0;4)&$(\frac14,\frac14,0)$&$(\frac14,\frac14-\Delta,-\Delta)$&27&(0,0,0;13)&(0,0,0;16)&$(\frac14,\frac14,0)$&$(\frac14,\frac14+\Delta,\Delta)$\\
4 &(0,0,0;2)&(0,0,0;3)&$(\frac14,-\frac14,0)$&$(\frac14,-\frac14-\Delta,\Delta)$&28&(0,0,0;14)&(0,0,0;15)&$(\frac14,-\frac14,0)$&$(\frac14,-\frac14+\Delta,-\Delta)$\\
5 &(0,0,0;2)&(0,0,0;4)&$(\frac14,0,-\frac14)$&$(\frac14+\Delta,-\Delta,-\frac14)$&29&(0,0,0;14)&(0,0,0;16)&$(\frac14,0,-\frac14)$&$(\frac14+\Delta,\Delta,-\frac14)$\\
6 &(0,0,0;2)&(0,0,0;5)&$(0,\frac14,\frac14)$&$(\Delta,\frac14+\Delta,\frac14)$&30&(0,0,0;15)&(0,0,0;16)&$(0,\frac14,-\frac14)$&$(\Delta,\frac14,-\frac14+\Delta)$\\
7 &(0,0,0;3)&(0,0,0;4)&$(0,\frac14,-\frac14)$&$(\Delta,\frac14,-\frac14-\Delta)$&31&(0,0,0;6)&(-1,1,0;11)&$(-\frac14,\frac14,0)$&$(-\frac14-\Delta,\frac14,\Delta)$\\
8 &(0,0,0;3)&(0,0,0;9)&$(\frac14,0,\frac14)$&$(\frac14,\Delta,\frac14+\Delta)$&32&(0,0,0;6)&(-1,0,1;16)&$(-\frac14,0,\frac14)$&$(-\frac14,-\Delta,\frac14+\Delta)$\\
9 &(0,0,0;4)&(0,0,0;13)&$(\frac14,\frac14,0)$&$(\frac14+\Delta,\frac14,\Delta)$&33&(0,0,0;6)&(0,1,1;1)&$(0,\frac14,\frac14)$&$(-\Delta,\frac14-\Delta,\frac14)$\\
10&(0,0,0;5)&(0,0,0;6)&$(0,\frac14,\frac14)$&$(\Delta,\frac14,\frac14+\Delta)$&34&(0,0,0;12)&(1,0,0;5)&$(\frac14,\frac14,0)$&$(\frac14+\Delta,\frac14,-\Delta)$\\
11&(0,0,0;5)&(0,0,0;7)&$(\frac14,0,\frac14)$&$(\frac14+\Delta,-\Delta,\frac14)$&35&(0,0,0;12)&(1,0,0;2)&$(\frac14,0,-\frac14)$&$(\frac14,-\Delta,-\frac14-\Delta)$\\
12&(0,0,0;5)&(0,0,0;8)&$(\frac14,\frac14,0)$&$(\frac14,\frac14-\Delta,\Delta)$&36&(0,0,0;15)&(1,0,0;5)&$(\frac14,0,\frac14)$&$(\frac14,\Delta,\frac14-\Delta)$\\
13&(0,0,0;6)&(0,0,0;7)&$(\frac14,-\frac14,0)$&$(\frac14,-\frac14-\Delta,-\Delta)$&37&(0,0,0;15)&(1,0,0;2)&$(\frac14,-\frac14,0)$&$(\frac14-\Delta,-\frac14,-\Delta)$\\
14&(0,0,0;6)&(0,0,0;8)&$(\frac14,0,-\frac14)$&$(\frac14-\Delta,-\Delta,-\frac14)$&38&(0,0,0;3)&(0,-1,0;8)&$(0,-\frac14,\frac14)$&$(\Delta,-\frac14+\Delta,\frac14)$\\
15&(0,0,0;7)&(0,0,0;8)&$(0,\frac14,-\frac14)$&$(-\Delta,\frac14,-\frac14+\Delta)$&39&(0,0,0;3)&(0,-1,0;14)&$(\frac14,-\frac14,0)$&$(\frac14+\Delta,-\frac14,\Delta)$\\
16&(0,0,0;7)&(0,0,0;10)&$(\frac14,-\frac14,0)$&$(\frac14-\Delta,-\frac14,\Delta)$&40&(0,0,0;9)&(0,-1,0;8)&$(-\frac14,-\frac14,0)$&$(-\frac14+\Delta,-\frac14,-\Delta)$\\
17&(0,0,0;8)&(0,0,0;14)&$(\frac14,0,-\frac14)$&$(\frac14,-\Delta,-\frac14+\Delta)$&41&(0,0,0;9)&(0,-1,0;14)&$(0,-\frac14,-\frac14)$&$(\Delta,-\frac14-\Delta,-\frac14)$\\
18&(0,0,0;9)&(0,0,0;10)&$(0,\frac14,\frac14)$&$(\Delta,\frac14,\frac14-\Delta)$&42&(0,0,0;4)&(0,0,-1;7)&$(0,\frac14,-\frac14)$&$(\Delta,\frac14+\Delta,-\frac14)$\\
19&(0,0,0;9)&(0,0,0;11)&$(\frac14,0,\frac14)$&$(\frac14+\Delta,\Delta,\frac14)$&43&(0,0,0;4)&(0,0,-1;10)&$(\frac14,0,-\frac14)$&$(\frac14,\Delta,-\frac14+\Delta)$\\
20&(0,0,0;9)&(0,0,0;12)&$(\frac14,\frac14,0)$&$(\frac14,\frac14+\Delta,-\Delta)$&44&(0,0,0;13)&(0,0,-1;7)&$(-\frac14,0,-\frac14)$&$(-\frac14,\Delta,-\frac14-\Delta)$\\
21&(0,0,0;10)&(0,0,0;11)&$(\frac14,-\frac14,0)$&$(\frac14,-\frac14+\Delta,\Delta)$&45&(0,0,0;13)&(0,0,-1;10)&$(0,-\frac14,-\frac14)$&$(-\Delta,-\frac14+\Delta,-\frac14)$\\
22&(0,0,0;10)&(0,0,0;12)&$(\frac14,0,-\frac14)$&$(\frac14-\Delta,\Delta,-\frac14)$&46&(0,0,0;1)&(-1,-1,0;16)&$(-\frac14,-\frac14,0)$&$(-\frac14+\Delta,-\frac14,\Delta)$\\
23&(0,0,0;11)&(0,0,0;12)&$(0,\frac14,-\frac14)$&$(-\Delta,\frac14,-\frac14-\Delta)$&47&(0,0,0;11)&(1,0,1;1)&$(\frac14,0,\frac14)$&$(\frac14,-\Delta,\frac14-\Delta)$\\
24&(0,0,0;12)&(0,0,0;15)&$(0,\frac14,-\frac14)$&$(\Delta,\frac14-\Delta,-\frac14)$&48&(0,0,0;16)&(0,1,-1;11)&$(0,\frac14,-\frac14)$&$(-\Delta,\frac14+\Delta,-\frac14)$\\
\hline\hline
\end{tabular}%
}
\end{table*}

\subsection{Unit-Cell Gauge and Dipole Moment}
\label{app:unit_cell_dipole}

Different assignments of sites and bonds to the reference unit cell
correspond to different unit cell gauges. Changing the unit-cell assignment
of a site or bond leaves the physical lattice unchanged but introduces a
phase factor in the corresponding row or column of $Q(\mathbf q)$. Consequently,
\(\det Q(\mathbf q)\) acquires an overall phase. Such a gauge transformation shifts the winding numbers by integers,
while leaving the zeros of \(\det Q(\mathbf q)\), their winding number
jumps, and total boundary mode counts unchanged
\cite{KaneLubensky2014,Lubensky_2015}.

This unit cell gauge can be characterized by assigning an auxiliary charge \(+d\) to each site and charge \(-1\) to each bond center
\cite{KaneLubensky2014,Lubensky_2015}. In three dimensions \(d=3\), the 16-site, 48-bond cubic unit cell is therefore neutral,
\(3\times16-48=0\), and the dipole moment of this charge distribution is
\[
    \mathbf R_L
    =
    3\sum_{\alpha=1}^{16}\mathbf r_\alpha
    -
    \sum_{\ell=1}^{48}\mathbf r_{\ell,\mathrm{mid}},
    \qquad
    \mathbf r_{\ell,\mathrm{mid}}
    =
    \frac{
    \mathbf R_{\ell1}+\mathbf r_{\alpha_{\ell1}}
    +\mathbf R_{\ell2}+\mathbf r_{\alpha_{\ell2}}
    }{2}.
\]
Using the site positions in Table~\ref{tab:Rstate_site_distortions} and
the bond information in Table~\ref{tab:bond_configuration}, we obtain
\[
    \mathbf R_L
    =
    (-1,1,1).
\]

This gauge dependence is directly visible in Fig.~\ref{fig:seven_panels}(a)--(c). For the
unit-cell convention used here, the 
\(\mathbf R_L=(-1,1,1)\) gives the background winding numbers \(-1\),
\(+1\), and \(+1\) for the \(x\), \(y\), and \(z\) directions,
respectively. A different unit-cell gauge would shift each figure uniformly
by an integer, but would not change the boundaries between regions with
different winding numbers. 

\subsection{\texorpdfstring{$P4_{3}32$}{P4_3 32} Symmetry
of the \texorpdfstring{$R$}{R}-State Induced Lattice}
\label{app:space_group}

In this section, we verify that the $R$ state is invariant under the space group
$P4_{3}32$ and the lattice distortion inherits this symmetry.

To express the site coordinates in the standard crystallographic setting of
$P4_{3}32$, we translate them by
\begin{equation}
\mathbf s
=
\left(
\frac18,\frac18,\frac18
\right).
\label{eq:P4332_origin_shift}
\end{equation}
Therefore the site coordinates are 
\begin{equation}
\widetilde{\mathbf r}_{i}
=
\mathbf r_i^{(0)}
+
\mathbf s.
\label{eq:shifted_basis_coordinates}
\end{equation}
We write the 24 space-group operations as
\begin{equation}
g_k
=
\left\{
R_k\middle|\mathbf t_k
\right\},
\qquad
\widetilde{\mathbf r}
\longmapsto
R_k\widetilde{\mathbf r}
+
\mathbf t_k,
\qquad
k=1,\ldots,24.
\label{eq:affine_space_group_operation}
\end{equation}
\begin{table*}[t]
\centering
\caption{The 24 symmetry operations of the space group $P4_{3}32$ and the
corresponding site permutations. Fixed sites are omitted from the cycle notation. Cycles
containing the spin-down sites $\{1,8,10,15\}$ are shown in blue.}
\label{tab:P4332_operations}
\label{tab:P4332_permutations}
\normalsize
\renewcommand{\arraystretch}{1.15}
\begin{tabular*}{0.92\textwidth}{@{\extracolsep{\fill}} l l}
\hline\hline
$g_k:(x,y,z)\mapsto(x',y',z')$ & $\pi_k$ \\
\hline
$(x,y,z)$ & $\mathrm{identity}$ \\
$\left(-x,y+\frac12,-z+\frac12\right)$ & $\blue{(1\,15)}(2\,16)(3\,13)(4\,14)(5\,11)(6\,12)(7\,9)\blue{(8\,10)}$ \\
$\left(x+\frac12,-y+\frac12,-z\right)$ & $\blue{(1\,10)}(2\,9)(3\,12)(4\,11)(5\,14)(6\,13)(7\,16)\blue{(8\,15)}$ \\
$\left(-x+\frac12,-y,z+\frac12\right)$ & $\blue{(1\,8)}(2\,7)(3\,6)(4\,5)(9\,16)\blue{(10\,15)}(11\,14)(12\,13)$ \\
$(z,x,y)$ & $(2\,3\,4)(5\,9\,13)(6\,11\,16)(7\,12\,14)\blue{(8\,10\,15)}$ \\
$(y,z,x)$ & $(2\,4\,3)(5\,13\,9)(6\,16\,11)(7\,14\,12)\blue{(8\,15\,10)}$ \\
$\left(z+\frac12,-x+\frac12,-y\right)$ & $\blue{(1\,10\,8)}(2\,12\,5)(3\,11\,7)(4\,9\,6)(13\,14\,16)$ \\
$\left(-z,x+\frac12,-y+\frac12\right)$ & $\blue{(1\,15\,10)}(2\,13\,11)(3\,14\,9)(4\,16\,12)(5\,7\,6)$ \\
$\left(-z+\frac12,-x,y+\frac12\right)$ & $\blue{(1\,8\,15)}(2\,6\,14)(3\,5\,16)(4\,7\,13)(9\,12\,11)$ \\
$\left(-y+\frac12,-z,x+\frac12\right)$ & $\blue{(1\,8\,10)}(2\,5\,12)(3\,7\,11)(4\,6\,9)(13\,16\,14)$ \\
$\left(y+\frac12,-z+\frac12,-x\right)$ & $\blue{(1\,10\,15)}(2\,11\,13)(3\,9\,14)(4\,12\,16)(5\,6\,7)$ \\
$\left(-y,z+\frac12,-x+\frac12\right)$ & $\blue{(1\,15\,8)}(2\,14\,6)(3\,16\,5)(4\,13\,7)(9\,11\,12)$ \\
$\left(-y+\frac14,-x+\frac14,-z+\frac14\right)$ & $(2\,11)(3\,6)(4\,16)(5\,9)(7\,14)\blue{(10\,15)}$ \\
$\left(-z+\frac14,-y+\frac14,-x+\frac14\right)$ & $(2\,16)(3\,11)(4\,6)(5\,13)\blue{(8\,10)}(12\,14)$ \\
$\left(-x+\frac14,-z+\frac14,-y+\frac14\right)$ & $(2\,6)(3\,16)(4\,11)(7\,12)\blue{(8\,15)}(9\,13)$ \\
$\left(y+\frac34,-x+\frac34,z+\frac14\right)$ & $\blue{(1\,15\,8\,10)}(2\,5\,7\,4)(3\,12\,6\,13)(9\,11\,16\,14)$ \\
$\left(-y+\frac34,x+\frac14,z+\frac34\right)$ & $\blue{(1\,10\,8\,15)}(2\,4\,7\,5)(3\,13\,6\,12)(9\,14\,16\,11)$ \\
$\left(y+\frac14,x+\frac34,-z+\frac34\right)$ & $\blue{(1\,8)}(2\,14)(4\,9)(5\,16)(7\,11)(12\,13)$ \\
$\left(z+\frac34,-y+\frac34,x+\frac14\right)$ & $\blue{(1\,15)}(3\,5)(4\,12)(6\,14)(7\,9)(11\,13)$ \\
$\left(z+\frac14,y+\frac34,-x+\frac34\right)$ & $\blue{(1\,8\,15\,10)}(2\,9\,16\,7)(3\,14\,13\,4)(5\,12\,11\,6)$ \\
$\left(-x+\frac34,z+\frac14,y+\frac34\right)$ & $\blue{(1\,10)}(2\,13)(3\,7)(5\,14)(6\,9)(12\,16)$ \\
$\left(x+\frac14,z+\frac34,-y+\frac34\right)$ & $\blue{(1\,8\,10\,15)}(2\,3\,9\,12)(4\,14\,11\,5)(6\,7\,13\,16)$ \\
$\left(-z+\frac34,y+\frac14,x+\frac34\right)$ & $\blue{(1\,10\,15\,8)}(2\,7\,16\,9)(3\,4\,13\,14)(5\,6\,11\,12)$ \\
$\left(x+\frac34,-z+\frac34,y+\frac14\right)$ & $\blue{(1\,15\,10\,8)}(2\,12\,9\,3)(4\,5\,11\,14)(6\,16\,13\,7)$ \\
\hline\hline
\end{tabular*}
\end{table*}
Each space group operation \(g_k\) induces a permutation \(\pi_k\) of the
sites, as listed in Table~\ref{tab:P4332_operations}. These permutations
preserve the spin assignment. Thus, the $R$ state is invariant under
$P4_{3}32$ symmetry.
The distortion is fixed locally by the $R$-state through Eq.~\eqref{SMeq:magnetopological_displacement} and therefore
introduces no additional symmetry breaking. 
The $R$-state induced topological
Maxwell lattice therefore retains the $P4_{3}32$ symmetry, which is also checked numerically.

\section{Equilibrium Matrix and Phonon Spectrum}
\label{app:phonon_spectrum}

For each oriented bond \(\ell\) in unit cell \(\mathbf R'\),
\(\widehat{\mathbf d}_{\ell}\) denotes the unit vector along the distorted
bond, and \(s_\ell(\mathbf R')\) denotes its stress. Let
\(\mathcal B(\mathbf R,\alpha)\) be the set of all bonds
connected to site \((\mathbf R,\alpha)\). The total force on this site is
\begin{equation}
\mathbf f_\alpha(\mathbf R)
=
\sum_{(\mathbf R',\ell)\in\mathcal B(\mathbf R,\alpha)}
\sigma_{\alpha\ell}\,
s_\ell(\mathbf R')\widehat{\mathbf d}_{\ell}.
\label{eq:app_force_real}
\end{equation}
Here \(\sigma_{\alpha\ell}=+1\) if bond \(\ell\) is oriented away from site
\((\mathbf R,\alpha)\), and \(-1\) if it is oriented toward the site.

For a periodic lattice, we use the Fourier transformations
\begin{equation}
\mathbf f_\alpha(\mathbf R)
=
\frac{1}{\sqrt N}\sum_{\mathbf q}
e^{i\mathbf q\cdot\mathbf R}\mathbf f_\alpha(\mathbf q),
\qquad
s_\ell(\mathbf R)
=
\frac{1}{\sqrt N}\sum_{\mathbf q}
e^{i\mathbf q\cdot\mathbf R}s_\ell(\mathbf q).
\label{eq:app_fourier}
\end{equation}
Substitution of Eq.~\eqref{eq:app_fourier} into
Eq.~\eqref{eq:app_force_real} gives
\begin{equation}
\mathbf f(\mathbf q)=Q(\mathbf q)\mathbf s(\mathbf q),
\label{eq:app_Qdef}
\end{equation}
where \(Q(\mathbf q)\) is the \(48\times48\) matrix, and the force
and stress vectors are
\begin{equation}
\mathbf f(\mathbf q)
=
\bigl(
f_1^x,f_1^y,f_1^z,\ldots,
f_{16}^x,f_{16}^y,f_{16}^z
\bigr)^T,
\qquad
\mathbf s(\mathbf q)
=
\bigl(s_1,\ldots,s_{48}\bigr)^T.
\label{eq:app_force_stress_vectors}
\end{equation}

The compatibility matrix describes the bond extensions induced by the
site displacements. Analogously, in momentum space, the displacement
components of the 16 sites and the extensions of the 48 bonds are
collected into
\begin{equation}
\mathbf u(\mathbf q)
=
\bigl(
u_1^x,u_1^y,u_1^z,\ldots,
u_{16}^x,u_{16}^y,u_{16}^z
\bigr)^T,
\qquad
\mathbf e(\mathbf q)
=
\bigl(e_1,\ldots,e_{48}\bigr)^T.
\label{eq:app_displacement_extension_vectors}
\end{equation}
The resulting
linear relation defines the compatibility matrix,
\begin{equation}
\mathbf e(\mathbf q)
=
C(\mathbf q)\mathbf u(\mathbf q).
\label{eq:app_Cdef}
\end{equation}
The compatibility matrix is the Hermitian
conjugate of the equilibrium matrix~\cite{Lubensky_2015},
\begin{equation}
C(\mathbf q)=Q^\dagger(\mathbf q).
\label{eq:app_CQ}
\end{equation}

In real space, the dynamical matrix is~\cite{Lubensky_2015}
\begin{equation}
D
=
\frac{1}{m}Q\mathcal K Q^T,
\label{eq:app_D_real_space}
\end{equation}
where \(\mathcal K\) is a diagonal matrix of bond stiffnesses and
\(m\) is the site mass.
For the periodic lattice considered here, Fourier transformation
block-diagonalizes this real-space dynamical matrix. Assuming a uniform bond
stiffness \(k_b\) and taking \(k_b/m\) as the unit, we obtain
\begin{equation}
D(\mathbf q)=Q(\mathbf q)Q^\dagger(\mathbf q).
\label{eq:app_Dq}
\end{equation}
The phonon spectrum is obtained from
\begin{equation}
D(\mathbf q)\bm{\epsilon}_n(\mathbf q)
=
\omega_n^2(\mathbf q)\bm{\epsilon}_n(\mathbf q),
\label{eq:app_eig}
\end{equation}
where $\omega_n(\mathbf q)$ is the frequency of the $n$th phonon mode
and $\bm{\epsilon}_n(\mathbf q)$ is the corresponding eigenvector.

\section{Cubic Symmetry of the Phonon Band Structure}
\label{app:phonon_symmetry}

To determine how the lattice symmetry constrains the phonon spectrum, we use
the standard treatment of lattice vibrations under space-group
operations~\cite{MaradudinVosko1968,DresselhausGroupTheory2008}. 

As shown in Sec.~\ref{app:space_group}, the $R$-state induced lattice is invariant under
$P4_{3}32$. We write a general element of this space group as
\begin{equation}
g=\{R_g|\mathbf t_g\},
\qquad
\mathbf x\longmapsto R_g\mathbf x+\mathbf t_g.
\label{eq:app_space_group_operation}
\end{equation}
For each site $\alpha$,
$\mathbf r_\alpha$ denotes its position within the reference unit cell.
Its image under $g$ can be written uniquely as
\begin{equation}
R_g\mathbf r_\alpha+\mathbf t_g
=
\mathbf r_{\pi_g(\alpha)}+\boldsymbol\Delta_{g\alpha},
\label{eq:app_basis_mapping}
\end{equation}
where $\pi_g$ is a permutation of the sixteen sites and
$\boldsymbol\Delta_{g\alpha}$ is a Bravais-lattice vector.  Hence the site label transforms as
\begin{equation}
(\mathbf R,\alpha)
\longmapsto
\bigl(R_g\mathbf R+\boldsymbol\Delta_{g\alpha},
\pi_g(\alpha)\bigr).
\label{eq:app_site_mapping}
\end{equation}

Then we derive the transformation law of the displacement field. Consider the
equilibrium and instantaneous positions of site
$(\mathbf R,\alpha)$,
\begin{equation}
\mathbf x_{\mathbf R\alpha}
=
\mathbf R+\mathbf r_\alpha,
\qquad
\widetilde{\mathbf x}_{\mathbf R\alpha}
=
\mathbf x_{\mathbf R\alpha}+\mathbf u_\alpha(\mathbf R).
\label{eq:app_equilibrium_instantaneous_positions}
\end{equation}
Applying $g$ to these two positions gives
\begin{equation}
\begin{aligned}
g\mathbf x_{\mathbf R\alpha}
&=
R_g\mathbf R+\mathbf r_{\pi_g(\alpha)}
+\boldsymbol\Delta_{g\alpha},\\
g\widetilde{\mathbf x}_{\mathbf R\alpha}
&=
R_g\mathbf R+\mathbf r_{\pi_g(\alpha)}
+\boldsymbol\Delta_{g\alpha}
+R_g\mathbf u_\alpha(\mathbf R).
\end{aligned}
\label{eq:app_transformed_positions}
\end{equation}
Because a displacement is a polar vector, the displacement field under
the action of the symmetry operation $g$, denoted by
$\mathbf u^{\,g}=g\mathbf u$, is the difference between the transformed
instantaneous and equilibrium positions. It therefore satisfies
\begin{equation}
\mathbf u^{\,g}_{\pi_g(\alpha)}
\bigl(R_g\mathbf R+\boldsymbol\Delta_{g\alpha}\bigr)
\equiv
g\widetilde{\mathbf x}_{\mathbf R\alpha}
-g\mathbf x_{\mathbf R\alpha}
=
R_g\mathbf u_\alpha(\mathbf R).
\label{eq:app_displacement_real_space_symmetry}
\end{equation}

To determine how the phonon spectrum transforms under
$g$, consider the $n$th mode at $\mathbf q$. At fixed $\mathbf q$,
Eq.~\eqref{eq:app_eig} gives 48 phonon modes labeled by
$n=1,\ldots,48$. Each eigenvector $\bm{\epsilon}_n(\mathbf q)$ has
48 components, grouped into 16 blocks containing the $x$, $y$, and $z$
displacement components of the individual sites;
$\bm{\epsilon}_{n,\alpha}(\mathbf q)$, with
$\alpha=1,\ldots,16$, denotes the block associated with site $\alpha$
in mode $n$. In the mode $n$, the displacement of site $\alpha$ in unit
cell $\mathbf R$ at time $t$ is
\begin{equation}
\mathbf u^{(n)}_\alpha(\mathbf R,t)
=
e^{i[\mathbf q\cdot\mathbf R-\omega_n(\mathbf q)t]}
\bm{\epsilon}_{n,\alpha}(\mathbf q).
\label{eq:app_phonon_normal_mode}
\end{equation}
Because $g$ is a symmetry of the lattice, applying it to this solution
produces another normal mode.
At the image site, the transformed mode can be written as
\begin{equation}
\bigl(\mathbf u^{(n)}\bigr)^{g}_{\pi_g(\alpha)}(\mathbf R',t)
=
e^{i[\mathbf q'\cdot\mathbf R'-\omega_{n'}(\mathbf q')t]}
\bm{\epsilon}^{\,g}_{n',\pi_g(\alpha)}(\mathbf q'),
\label{eq:app_transformed_normal_mode}
\end{equation}
where $\mathbf q'$ and $n'$ label the transformed mode. The image-cell
label is
\begin{equation}
\mathbf R'
=
R_g\mathbf R+\boldsymbol\Delta_{g\alpha}.
\label{eq:app_image_cell}
\end{equation}
Using Eqs.~\eqref{eq:app_displacement_real_space_symmetry} and
\eqref{eq:app_phonon_normal_mode} gives
\begin{equation}
\bigl(\mathbf u^{(n)}\bigr)^{g}_{\pi_g(\alpha)}(\mathbf R',t)
=
e^{i[\mathbf q\cdot\mathbf R-\omega_n(\mathbf q)t]}
R_g\bm{\epsilon}_{n,\alpha}(\mathbf q).
\label{eq:app_transformed_mode_intermediate}
\end{equation}
Since
$\mathbf R=R_g^T(\mathbf R'-\boldsymbol\Delta_{g\alpha})$, this becomes
\begin{equation}
\bigl(\mathbf u^{(n)}\bigr)^{g}_{\pi_g(\alpha)}(\mathbf R',t)
=
e^{i[(R_g\mathbf q)\cdot\mathbf R'-\omega_n(\mathbf q)t]}
e^{-i(R_g\mathbf q)\cdot\boldsymbol\Delta_{g\alpha}}
R_g\bm{\epsilon}_{n,\alpha}(\mathbf q).
\label{eq:app_transformed_mode_rewritten}
\end{equation}
Comparison with Eq.~\eqref{eq:app_transformed_normal_mode} gives
\begin{equation}
\begin{gathered}
\mathbf q'=R_g\mathbf q,
\qquad
\omega_{n'}(R_g\mathbf q)=\omega_n(\mathbf q),\\
\bm{\epsilon}^{\,g}_{n',\pi_g(\alpha)}(R_g\mathbf q)
=
e^{-i(R_g\mathbf q)\cdot\boldsymbol\Delta_{g\alpha}}
R_g\bm{\epsilon}_{n,\alpha}(\mathbf q).
\end{gathered}
\label{eq:app_displacement_momentum_symmetry}
\end{equation}
Since every space-group operation is invertible,
Eq.~\eqref{eq:app_displacement_momentum_symmetry} defines a one-to-one
correspondence between the phonon modes at \(\mathbf q\) and
\(R_g\mathbf q\). Therefore, the phonon spectra at \(\mathbf q\) and
\(R_g\mathbf q\) are identical,
\begin{equation}
\left\{\omega_n(R_g\mathbf q)\right\}_{n=1}^{48}
=
\left\{\omega_n(\mathbf q)\right\}_{n=1}^{48}.
\label{eq:app_omega_space_group}
\end{equation}

There is a second constraint that follows from the reality of the
real-space matrix elements.  The bond directions and force constants are
real.  Hence
\begin{equation}
Q(-\mathbf q)=Q^*(\mathbf q),
\qquad
D(-\mathbf q)=D^*(\mathbf q).
\label{eq:app_D_reality}
\end{equation}
Taking the complex conjugate of
Eq.~\eqref{eq:app_eig} then gives
\begin{equation}
\omega_n(-\mathbf q)=\omega_n(\mathbf q),
\qquad
\bm{\epsilon}_n(-\mathbf q)
\sim
\bm{\epsilon}_n^*(\mathbf q),
\label{eq:app_phonon_reality}
\end{equation}

The rotational part of $P4_{3}32$, together with
$\omega_n(-\mathbf q)=\omega_n(\mathbf q)$, give
\begin{equation}
\left\{\omega_n(\pm R_g\mathbf q)\right\}_{n=1}^{48}
=
\left\{\omega_n(\mathbf q)\right\}_{n=1}^{48}.
\label{eq:app_full_octahedral_spectrum}
\end{equation}
Thus the phonon spectrum has cubic symmetry in momentum
space.  In particular,
the $x$, $y$, and $z$ directions are symmetry equivalent.  This explains
why the winding number maps calculated along the three directions in
Fig.~\ref{fig:seven_panels}(a)--(c) have the same pattern.

\section{Zero-Mode Lines in Three-Dimensional Maxwell Lattices}
\label{app:weyl_line_codimension}

To see why bulk zero modes generically form lines in a 3D Maxwell lattice,
consider a general two-band Hamiltonian,
\begin{equation}
H({\bf q})
=
d_0({\bf q})\sigma_0
+d_x({\bf q})\sigma_x
+d_y({\bf q})\sigma_y
+d_z({\bf q})\sigma_z ,
\end{equation}
where all four coefficients are real. Its two eigenvalues are
\begin{equation}
E_{\pm}({\bf q})
=
d_0({\bf q})
\pm
\sqrt{d_x^2({\bf q})+d_y^2({\bf q})+d_z^2({\bf q})}.
\end{equation}
A band degeneracy therefore requires
\begin{equation}
d_x({\bf q})=d_y({\bf q})=d_z({\bf q})=0 .
\end{equation}
These are three independent real conditions. Their solutions in a
3D Brillouin zone are thus generically isolated points,
corresponding to Weyl points.

The codimension is reduced in the presence of chiral symmetry,
\begin{equation}
\Gamma H({\bf q})\Gamma^{-1}=-H({\bf q}) .
\end{equation}
Taking $\Gamma=\sigma_z$ forbids both the $\sigma_0$ and $\sigma_z$
terms, leaving
\begin{equation}
H({\bf q})=d_x({\bf q})\sigma_x+d_y({\bf q})\sigma_y .
\end{equation}
A zero-energy degeneracy now requires only
\begin{equation}
d_x({\bf q})=d_y({\bf q})=0 .
\end{equation}
The resulting zero modes generically form lines in three dimensions.

The same structure arises in topological mechanics. The
effective Hamiltonian is~\cite{KaneLubensky2014}
\begin{equation}
\mathcal H_{\rm KL}({\bf q})
=
\begin{pmatrix}
0 & Q({\bf q}) \\
Q^\dagger({\bf q}) & 0
\end{pmatrix},
\label{eq:app_KL_hamiltonian}
\end{equation}
which obeys
\begin{equation}
\left\{\Gamma,\mathcal H_{\rm KL}({\bf q})\right\}=0,
\qquad
\Gamma
=
\begin{pmatrix}
I & 0 \\
0 & -I
\end{pmatrix}.
\end{equation}

The same reduction in codimension can be seen directly from the
equilibrium matrix. For a Maxwell lattice, $Q({\bf q})$ is square, and
the zero-frequency condition is simply
\begin{equation}
\det Q^{\dagger}({\bf q})=0 .
\label{eq:app_detQ_zero}
\end{equation}
At a generic momentum, $\det Q^{\dagger}({\bf q})$ is complex, so the zero-mode
condition imposes two independent real constraints,
\begin{equation}
\mathrm{Re}\det Q^{\dagger}({\bf q})=0,
\qquad
\mathrm{Im}\det Q^{\dagger}({\bf q})=0 .
\end{equation}
Thus, non-Goldstone zero modes generically form lines in
3D Maxwell lattices.

Each Weyl line carries an integer winding number evaluated along a
small closed loop $\mathcal C$ that links the line,
\begin{equation}
n_W
=
\frac{1}{2\pi i}
\oint_{\mathcal C}
d{\bf q}\cdot\nabla_{\bf q}\ln\det Q({\bf q}) .
\label{eq:app_Weyl_line_charge}
\end{equation}
A nonzero $n_W$ prevents the line from being removed by a small
perturbation.
\section{Boundary Zero Modes}
\label{app:boundary_mode}

\subsection{Counting Boundary Zero Modes}

For boundaries normal to \(\mathbf a_x\), the complex inverse penetration depths \(\kappa\) at each
boundary momentum \(\mathbf q_\parallel=q_y\mathbf b_y+q_z\mathbf b_z\) are obtained from 
~\cite{KaneLubensky2014,Lubensky_2015}
\begin{equation}
    \det C(i\kappa,q_y,q_z)=0 .
    \label{eq:app_complex_surface_condition}
\end{equation}
Since \(C(\mathbf q)=Q^\dagger(\mathbf q)\) [Eq.~\eqref{eq:app_CQ}],
this is equivalent to the condition $\det Q^{\dagger}(i\kappa,q_y,q_z)=0 $ in
the main text. With the convention used here,
the solutions with \(\operatorname{Re}\kappa>0\) and
\(\operatorname{Re}\kappa<0\) correspond to
boundary modes localized on the boundaries with outward normals
\(-\mathbf a_x\) and \(+\mathbf a_x\), respectively. The numbers of solutions give the corresponding numbers of
boundary zero modes per boundary unit cell.

\begin{figure}[!htbp]
    \centering
    \begin{minipage}[t]{0.46\textwidth}
        \centering
        \begin{overpic}[height=0.34\textheight]{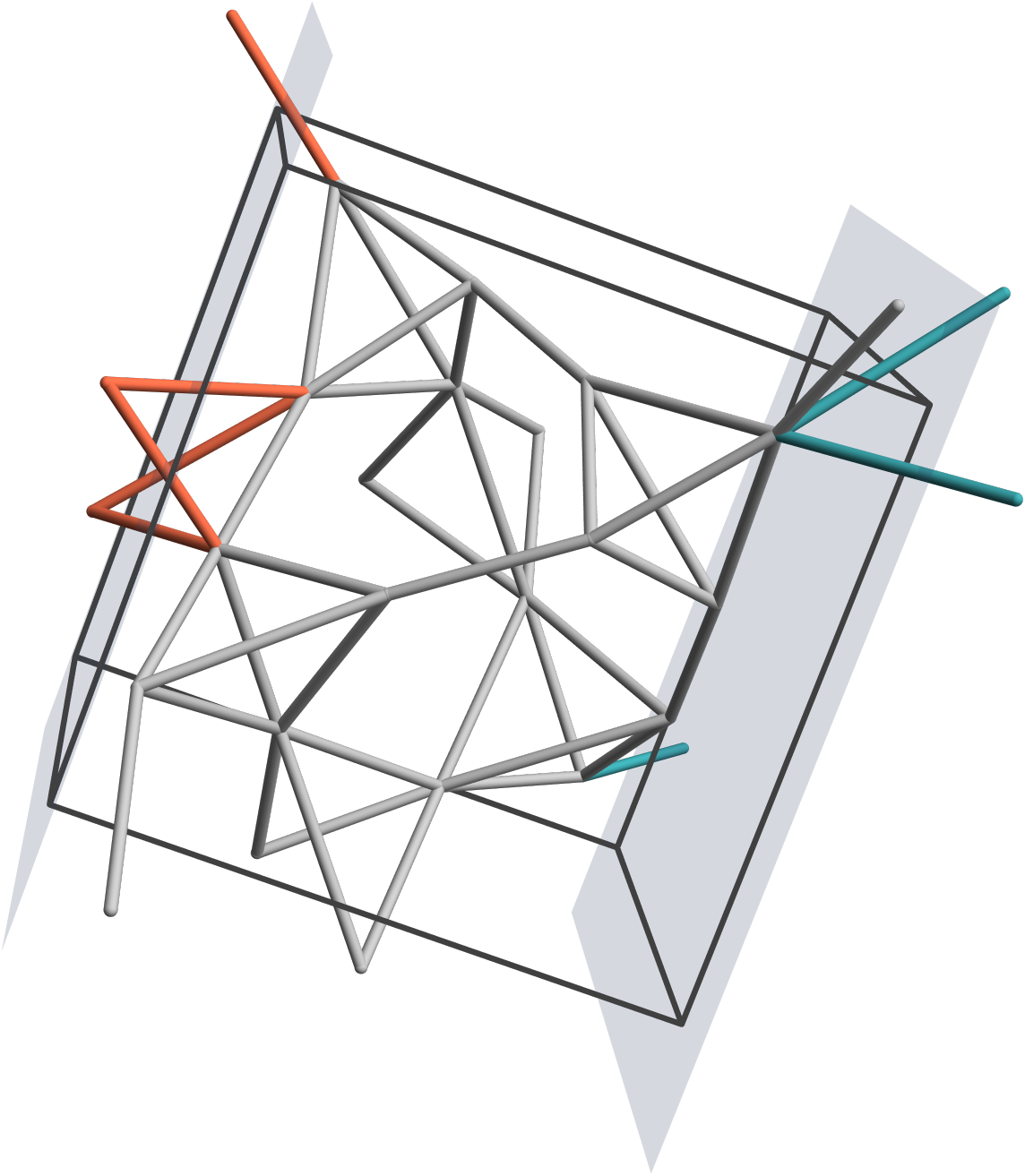}
            \put(2,94){\makebox(0,0)[lb]{(a)}}
        \end{overpic}
    \end{minipage}
    \hspace{0.03\textwidth}
    \begin{minipage}[t]{0.46\textwidth}
        \centering
        \begin{overpic}[height=0.34\textheight]{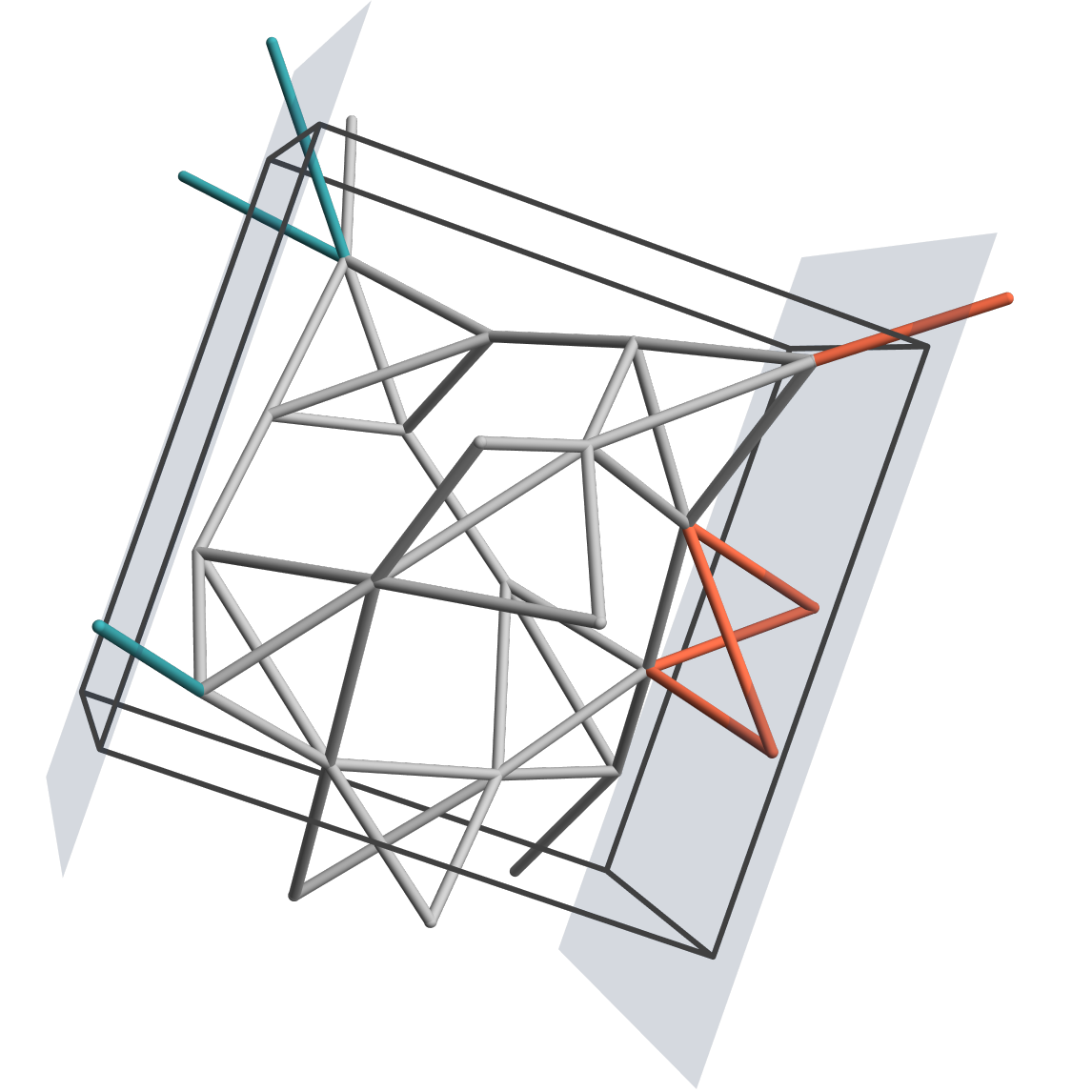}
            \put(2,94){\makebox(0,0)[lb]{(b)}}
        \end{overpic}
    \end{minipage}
    \caption{
    Boundary terminations normal to \(\mathbf a_x\). Three green bonds are
    cut at the boundary with outward normal \(-\mathbf a_x\), and five orange
    bonds at the boundary with outward normal \(+\mathbf a_x\).
    }
    \label{fig:app_surface_zero_modes}
\end{figure}

At a fixed \(\mathbf q_\parallel\), the number of boundary zero modes per boundary
unit cell separates into local and topological contributions~\cite{KaneLubensky2014},
\begin{equation}
    \nu(\mathbf q_\parallel,\mathbf G)
    =
    \nu_L(\mathbf G)
    +
    \nu_T(\mathbf q_\parallel,\mathbf G).
\end{equation}
Here \(\mathbf G=+\mathbf b_x\) and \(-\mathbf b_x\) correspond to the boundaries
whose outward normals point along \(+\mathbf a_x\) and \(-\mathbf a_x\),
respectively.
Here \(\nu_L\) depends only on the boundary termination, whereas the
topological contribution is determined by the bulk winding numbers,
\begin{equation}
    \nu_T(\mathbf q_\parallel,\mathbf G)
    =
    \frac{\mathbf G}{2\pi}\cdot
    \mathbf R_T(\mathbf q_\parallel),
    \qquad
    \mathbf R_T(\mathbf q_\parallel)
    =
    \sum_i n_i(\mathbf q_\parallel)\mathbf a_i .
\end{equation}
Here \(n_i(\mathbf q_\parallel)\) are the winding numbers discussed in the main text.

Figure~\ref{fig:app_surface_zero_modes} illustrates the local contribution
for the two boundaries whose normals are parallel to \(\pm \mathbf a_x\). Within one boundary
unit cell, following the convention introduced in
Sec.~\ref{app:Rstate_geometry}, the three bonds
\(\ell=31,32,46\) (green bonds) are cut at the
boundary with outward normal \(-\mathbf a_x\), whereas the five bonds
\(\ell=34,35,36,37,47\) (orange bonds) are cut at the boundary with
outward normal \(+\mathbf a_x\). Their local zero-mode counts are
therefore three and five, respectively. When \(\nu_T=0\), this agrees with the eight solutions
of \(\det C(i\kappa,q_y,q_z)=0\) discussed in the main text: three have
\(\operatorname{Re}\kappa>0\), while the other five have
\(\operatorname{Re}\kappa<0\). As \(\mathbf q_\parallel\) crosses a
projected Weyl line, the resulting change in \(\nu_T\) redistributes the
modes between the two boundaries.

\subsection{Inverse Penetration Depth}

For boundaries normal to \(\mathbf a_x\), the complex momentum
is
\begin{equation}
    \mathbf q=(i\kappa,q_y,q_z).
\end{equation}
For each fixed boundary momentum, directly solving
\(\det C(i\kappa,q_y,q_z)=0\) for \(\kappa\) requires finding the roots
of a high-order nonlinear equation and is numerically impractical.
To avoid this problem, we introduce
$\lambda=e^\kappa$ to recast it as a quadratic
eigenvalue problem.

The momentum dependence of \(Q(\mathbf q)\) follows from the real-space
force relation in Eq.~\eqref{eq:app_force_real} and the Fourier transform
in Eq.~\eqref{eq:app_fourier}. Specifically, for a bond \(\ell\) connected
to site \((\mathbf R,\alpha)\) and assigned to unit cell
\(\mathbf R+\Delta\mathbf R\), the Fourier-transformed stress is
\[
    s_\ell(\mathbf R+\Delta\mathbf R)
    =
    \frac{1}{\sqrt N}\sum_{\mathbf q}
    e^{i\mathbf q\cdot\mathbf R}
    e^{i\mathbf q\cdot\Delta\mathbf R}
    s_\ell(\mathbf q).
\]
The contribution of this bond to \(Q(\mathbf q)\) carries the intercell
factor \(e^{i\mathbf q\cdot\Delta\mathbf R}\). With
\(\mathbf q=(i\kappa,q_y,q_z)\) and \(\lambda=e^\kappa\), this factor becomes
\begin{equation}
    e^{i\mathbf q\cdot\Delta\mathbf R}
    =
    e^{i(q_y\Delta R_y+q_z\Delta R_z)}
    \lambda^{-\Delta R_x}.
\end{equation}
Since bonds only connect the same or neighboring unit cells along the
\(x\) direction, \(\Delta R_x\in\{-1,0,+1\}\), so \(Q(\mathbf q)\)
at fixed boundary momentum \((q_y,q_z)\) can be written as
\begin{equation}
    Q(\lambda)
    =
    \lambda^{-1} Q_{-1}
    +
    Q_0
    +
    \lambda Q_{+1}.
\end{equation}
The matrices \(Q_{-1}\), \(Q_0\), and \(Q_{+1}\) collect the terms
proportional to \(\lambda^{-1}\), \(1\), and \(\lambda\), respectively.
Using \(C=Q^\dagger\), the compatibility matrix can be written as
\begin{equation}
    C(\lambda)
    =
    \lambda^{-1} C_{-1}
    +
    C_0
    +
    \lambda C_{+1}.
\end{equation}
Here \(C_{-1}=Q_{+1}^\dagger\), \(C_0=Q_0^\dagger\), and
\(C_{+1}=Q_{-1}^\dagger\).

The zero-mode condition \(\det C(\lambda)=0\) is equivalent to
\begin{equation}
    C(\lambda)\mathbf u=0 .
\end{equation}
Multiplication by \(\lambda\) yields the quadratic matrix-polynomial
problem
\begin{equation}
    \left(
    C_{-1}
    +
    \lambda C_0
    +
    \lambda^2 C_{+1}
    \right)\mathbf u=0 .
\end{equation}
To apply standard generalized-eigenvalue solvers, this quadratic matrix
polynomial is converted to its first companion linearization
\cite{doi:10.1137/S0036144500381988}. Introducing
\begin{equation}
    \mathbf v=
    \begin{pmatrix}
        \mathbf u\\
        \lambda\mathbf u
    \end{pmatrix},
\end{equation}
gives the generalized eigenvalue equation
\begin{equation}
    \begin{pmatrix}
        0&I\\
        -C_{-1}&-C_0
    \end{pmatrix}
    \mathbf v
    =
    \lambda
    \begin{pmatrix}
        I&0\\
        0&C_{+1}
    \end{pmatrix}
    \mathbf v .
\end{equation}
Here \(I\) is the \(48\times48\) identity matrix. Denoting the two
\(96\times96\) block matrices on the left- and right-hand sides by
\(\mathcal A\) and \(\mathcal B\), respectively, the linearized problem
takes the matrix-pencil form
\(\mathcal L(\lambda)\mathbf v=0\), with
\(\mathcal L(\lambda)=\mathcal A-\lambda\mathcal B\). For each fixed
\((q_y,q_z)\), the generalized eigenvalues \(\lambda_i\) are computed
using Mathematica. Eigenvalues with \(|\lambda_i|\leq10^{-6}\) or
\(|\lambda_i|\geq10^{6}\) are treated as zero or infinite, respectively,
and discarded. For each remaining finite, nonzero eigenvalue, the relation
\(\lambda_i=e^{\kappa_i}\) gives
\begin{equation}
    \operatorname{Re}\kappa_i=\log|\lambda_i| .
\end{equation}

\section{Robustness of the Weyl-Line Topology}
\label{app:nonuniform_stiffness}
The phonon calculation in the main text assumes the same stiffness for all
bonds. In a real magnetic material, higher-order terms in the bond-length
expansion of $J(r)$ and other microscopic interactions modify the elastic
response. Within the harmonic model used here, these effects are incorporated
into effective bond-stretching stiffnesses.

For the $R$-state induced equilibrium lattice, these stiffnesses are
collected into the diagonal matrix
\begin{equation}
    \mathcal K
    =
    \operatorname{diag}
    \left(
    \mathcal K_1,
    \mathcal K_2,
    \ldots,
    \mathcal K_{N_{\mathrm{bond}}}
    \right).
\label{eq:app_general_stiffness_matrix}
\end{equation}
Assuming that the renormalization does not drive any bond stiffness to
zero or negative values, the
matrix \(\mathcal K\) remains positive definite. Its square root is
therefore well defined and invertible, and
Eq.~\eqref{eq:app_D_real_space} can be written as
\begin{equation}
    D
    =
    \frac{1}{m}\widetilde Q\widetilde Q^T,
    \qquad
    \widetilde Q=Q\mathcal K^{1/2}.
\label{eq:app_weighted_dynamical_matrix}
\end{equation}
The positive definiteness of \(\mathcal K\) then implies
\begin{equation}
   \dim \ker\widetilde Q^T=\dim \ker Q^T.
\label{eq:app_kernel_invariance}
\end{equation}
Thus, the stiffness renormalization changes the nonzero phonon
frequencies but leaves the number of zero modes unchanged.

The argument above applies to general positive bond-stiffness
renormalizations and does not rely on periodicity. In the model considered
here, the dominant renormalization arises from the SLC and therefore inherits
the periodicity of the \(R\)-state unit cell. This periodicity allows us to
Fourier transform the lattice equations and extend the argument to the
momentum-space topology. The 48 stiffnesses within one unit cell are collected in the
diagonal matrix
\begin{equation}
    \mathcal K_{\mathrm{eff}}
    =
    \operatorname{diag}
    \left(
    \mathcal K_1^{\mathrm{eff}},\mathcal K_2^{\mathrm{eff}},
    \ldots,\mathcal K_{48}^{\mathrm{eff}}
    \right).
\end{equation}
Then the dynamical matrix in momentum space is 
\begin{equation}
    D_{\mathrm{eff}}(\mathbf q)
    =
    Q(\mathbf q)\mathcal K_{\mathrm{eff}}Q^\dagger(\mathbf q).
\label{eq:app_D_nonuniform_K}
\end{equation}
The stiffness-weighted equilibrium matrix is
\begin{equation}
    \widetilde Q(\mathbf q)
    =
    Q(\mathbf q)\mathcal K_{\mathrm{eff}}^{1/2},
\end{equation}
and its determinant is
\begin{equation}
    \det\widetilde Q(\mathbf q)
    =
    \det Q(\mathbf q)\sqrt{\det\mathcal K_{\mathrm{eff}}}.
\end{equation}
Because \(\sqrt{\det\mathcal K_{\mathrm{eff}}}\) is real, positive, and
independent of \(\mathbf q\), it changes neither the zeros nor the phase
winding of \(\det Q(\mathbf q)\). Thus, the Weyl-line zero modes and their
winding numbers obtained in the uniform central-force model remain unchanged
under periodic renormalizations that keep all bond stiffnesses positive.

\end{document}